\documentclass[prd,twocolumn,showpacs,amsmath,amssymb,superscriptaddress]{revtex4}
\usepackage{graphicx}
\usepackage{dcolumn}
\usepackage{bm}
\usepackage{color}

\begin{document}
\title{Perturbational Treatment of the Gravitational Potential Effect on Binary
Black Hole Evolution}

\author{Zhoujian Cao\footnote{zjcao@amt.ac.cn}}
\affiliation{Institute of Applied Mathematics, Academy of
Mathematics and Systems Science, Chinese Academy of Sciences,
Beijing 100190, China}

\author{Jui-Ping Yu}
\author{Chun-Yu Lin}
\affiliation{Department of Physics and National Center for Theoretical Sciences, National Cheng-Kung University, Tainan 701, Taiwan}

\author{Shan Bai}
\affiliation{Institut de Math\'ematiques de Bourgogne, Universit\'e de
Bourgogne, 9 avenue Alain Savary, 21078 Dijon Cedex, France}

\author{Hwei-Jang Yo\footnote{hjyo@phys.ncku.edu.tw}}
\affiliation{Department of Physics, National Cheng-Kung University, Tainan 701,
Taiwan}

\begin{abstract}
Binary black hole (BBH) systems are usually located in the gravitational
potential well formed by a massive black hole (BH), which is mostly
located in the center of a galaxy. In most existing studies, the BBH
systems are treated as isolated systems, while the effect of the
background is ignored. The validity of the approximation is based on
the belief that the background gravitational field from other
sources is extremely weak compared with the strong gravitational
field produced by the BBH itself during the evolution, and can be
neglected in gravitational wave detection. However, it is still
interesting to check how valid this approximation is. In this work,
instead of simulating the three-BH problem with a fully relativistic
treatment, we use a perturbational scheme to investigate the effect
of the background gravitational potential on the evolution of a BBH,
especially on the waveform of its gravitational radiation. Four
scenarios are considered including the head-on collision and the
inspiral-to-merger process of a BBH which is either freefalling
towards or circularly orbiting around a third large BH. The head-on
collision and the circular inspiral are two limits of all
possible configurations. The existence of the background
gravitational potential changes the arrival time of the
gravitational wavefront of a BBH, prolongs the wavelength, and
increases the gravitational radiation energy. And most
interestingly, the background gravitational potential induces the
higher-order modes of the gravitational wave of a BBH. These interesting
phenomena can be explained by the gravitational redshift effect and
the change of eccentricity of a BBH's orbit from the background gravitational
potential. Without further studies, these phenomena could introduce
complications or even mislead people in the identification of the source of
gravitational wave and in distinguishing the signatures of an
isolated BBH from a BBH in a background gravitational potential.
\end{abstract}
\date{\today}

\maketitle
%%%%%%%%%%%%%%%%%%%%%%%%%%%%%%%%%%%%%%%%%%%%%%%%%%%%%%%%%%%%%%%%%%
\section{Introduction}
%%%%%%%%%%%%%%%%%%%%%%%%%%%%%%%%%%%%%%%%%%%%%%%%%%%%%%%%%%%%%%%%%%
In numerical relativity, Einstein's equation can be solved numerically without
any approximation or symmetry assumption, with the aid of a supercomputer.
Besides the massive computational cost involved,
the stability issue is also a nontrivial problem in the numerical calculations
for Einstein's equation.  Breakthroughs in 2005 and 2006
\cite{bbhsuccess,moving_punctur1,moving_punctur2}
shifted the development of numerical relativity to a higher pace.
Now many numerical relativity groups around the world are capable of evolving
black hole (BH) systems with high accuracy.
Moreover, a variety of interesting topics, such as long term gravitational
waves \cite{frans07,baker07a,boyle07,vaishnav07,baker07b,ajith07,buonanno07,
baumgarte08,pan08,husa08,hannam08}, gravitational radiation induced BH
recoil \cite{krishnan07,baker06,sopuerta06,gonzalez07a,sopuerta07,
herrmann06,herrmann07a,campanelli07a,koppitz07,choi07,gonzalez07b,
baker07c,campanelli07b,berti07,tichy07,herrmann07b}, the estimation of the
final BH's mass and the spin of binary black hole systems
\cite{campanelli07c,boyle08,tichy08,marronetti08,rezzolla09,washik08},
and so on
\cite{lousto08a,lousto08b,sperhake08,campanelli09,meter09,lovelace09},
have been extensively studied over the past few years.
Numerical relativity has now become an efficient tool in the research of
general relativity and astrophysics.

Gravitational wave detection is an important task for the further
development of gravitation and general relativity.
Currently, the ground-based laser interferometers such as LIGO \cite{LIGO},
VIRGO \cite{VIRGO}, GEO600 \cite{GEO}, and TAMA \cite{TAMA} are already up and
running, while the space-based laser interferometer LISA \cite{LISA} is still
under construction.
Progress in detection has made the requirement for the theoretical prediction
of gravitational wave signal more and more urgent.
Numerical relativity plays a key role in both aiding gravitational wave data
analysis \cite{ninja} and helping to build template banks of the
phenomenological waveform \cite{phenomenological} for the most important
 gravitational source---binary black hole (BBH) coalescence.

In most theoretical and numerical calculations, BBHs are treated as isolated
systems, although stellar-mass BBHs usually are located in a galaxy with a
super-massive BH in the center.
The validity of the approximation is based on the idea that a
stellar-mass BBH is usually far away from the super-massive BH whose
gravitational potential is negligible compared with the strength of
the gravitational field produced by the BBH itself. Besides the
possible super-massive-BH environment, it is possible for
stellar-mass BBHs to be affected by other stellar BHs and/or stars,
especially in globular clusters. Although the background potentials
in these scenarios are usually much weaker than the gravitational
field of the BBH, it still remains as an interesting question as to
how valid the approximation is, and to what extent the approximation
holds. There have been several studies of the three-BH problem with
numerical relativity from various standpoints
\cite{campanelli06,lousto08a,lousto08b,LCNH08,GPBC10,threebs}. The
results all indicate that the dynamics and behavior displayed by
three BHs are qualitatively different from those of an isolated BBH. For
example, in \cite{campanelli06}, numerically generated BBH initial
data sets and post-Newtonian techniques are utilized to show that
the presence of a third BH has non-negligible relativistic effects
on the location of a BBH's innermost stable circular orbit (ISCO),
an increase in merging time, and an amplification of the
gravitational radiation emitted from the BBH. As we all know,
information about a BBH such as the initial/final spin, the mass ratio
and the orientation can be extracted from a detected gravitational wave
by comparing the wave train with the waveform template in the data
bank calculated with theoretical models. Without proper
consideration of the environmental effect on the BBH evolution and
thus on the distortion of the gravitational wave, the extracted
information might be incorrect and possibly lead to
misinterpretation and/or misunderstanding. It is also possible that
the nonlinearity of general relativity might give rise to a
qualitatively difference due to a tiny difference, just like in
other nonlinear systems \cite{jhingan01}. It is also interesting to
seek any possible nonlinear effects in the three-BH problem which
might be helpful in distinguishing among different scenarios
\cite{torigoe09}.

In this work, instead of the heavily numerical calculation of the evolution of
a three-BH problem with a fully relativistic treatment, we use a perturbational
scheme, along with numerical simulations, to investigate the effect of the
background gravitational potential on the evolution of a BBH, especially on
the waveform of the gravitational radiation. 
As with our previously preliminary study \cite{old3BH}, the scenarios we considered
include the head-on collision and the
inspiral-to-merger process of a BBH, considered as the two limits in
all possible configurations, in the cases where the BBH is freefalling towards
or circularly orbiting around the third large BH.
Our results show that the existence of the background gravitational potential
changes the arrival time of the gravitational wavefront of the BBH,
prolongs the wavelength, and increases the gravitational radiation
energy. And most interestingly, the background gravitational
potential induces the higher-order modes of the gravitational wave of a BBH.
These interesting phenomena can be explained by the gravitational
redshift effect and the change of eccentricity of the BBH's orbit from
the gravitational potential. These phenomena could introduce complications or
even mislead people in the identification of the source of
gravitational wave without further studies to distinguish the
signatures of a BBH with a background gravitational potential from
the ones of an isolated BBH.
%%%%%%%%%%%%%%%%%%%%%%%%%%%%%%%%%%%%%%%%%%%%%%%%%%%%%%%%%%%%%%%
\begin{figure}[tbp]
\centering\includegraphics[width=\columnwidth]{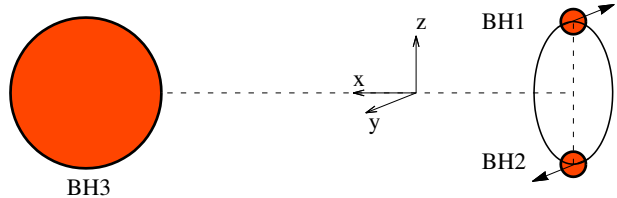}\caption{A schematic illustration of a BBH system in the gravitational
potential of a third massive BH.}
\label{fig1}
\end{figure}
%%%%%%%%%%%%%%%%%%%%%%%%%%%%%%%%%%%%%%%%%%%%%%%%%%%%%%%%%%%%%%%

The remainder of the paper is organized as follows:
In Sec.~\ref{secii}, we give a description of the code configuration, the
gauge condition, the initial data, and especially the perturbational method for 
the outer boundary condition.
Then we consider the four special cases for the evolution of a BBH under the
influence of a third large BH, followed by a numerical analysis of
these results one by one in Sec.~\ref{seciii}.
We summarize and discuss the viability of the perturbational method
and the implications of our findings in Sec.~\ref{seciv}.
Throughout the paper, the geometric units with $G=c=1$ are used.
%%%%%%%%%%%%%%%%%%%%%%%%%%%%%%%%%%%%%%%%%%%%%%%%%%%%%%%%%%%%%%%%%%%
\section{Numerical techniques} \label{secii}
%%%%%%%%%%%%%%%%%%%%%%%%%%%%%%%%%%%%%%%%%%%%%%%%%%%%%%%%%%%%%%%
As a preliminary study of the background effect, we investigate the dynamics
and the waveforms of a BBH system in a potential background which comes from
a third, distant and massive BH.
For this three-body system, we do not simulate its three-body dynamics with
a fully relativistic treatment as in \cite{threebs}.
Instead, a method of perturbation is adopted for the problem since the
gravitational field from the third BH to the BBH is weaker than the
field between the two small BHs in the BBH.
It is obvious that the effect of the third massive BH is stronger if it has
larger mass $m_3$ or the shorter distance $R$ to the binary system, or both.
Therefore, we would use the ratio of $m_3/R$ to denote the strength of the
gravitational background resulting from the third massive BH.
Figure \ref{fig1} illustrates the configuration considered in this work.
%%%%%%%%%%%%%%%%%%%%%%%%%%%%%%%%%%%%%%%%%%%%%%%%%%%%%%%%%%%%%%%
\subsection{Code configuration and Gauge condition}
%%%%%%%%%%%%%%%%%%%%%%%%%%%%%%%%%%%%%%%%%%%%%%%%%%%%%%%%%%%%%%%
Our AMSS-NCKU code for solving Einstein's field equations based on
the BSSN formalism is updated from our previous work \cite{cao08}.
For the time integration, we updated from the iterative second-order
Crank-Nicholson method to the fourth-order Runge-Kutta method. We also
developed our own infrastructure for the implementation of the fixed
mesh refinement and its parallelization in addition to using the
GrACE package \cite{grace}. Our new infrastructure also provides a
``moving box'' style mesh refinement. The box will move to follow
the position of the BHs, which are determined by \cite{moving_punctur1}
\begin{equation}
\frac{\rm d}{{\rm d}t}x^i_{BH}=-\beta^i(x^i_{BH}).
\end{equation}
We follow closely \cite{brugmann08,shibata08} in the development of the new
infrastructure to deal with the interface between two mesh levels.
Since three time levels are used for the interpolation in time, we adopt the
scheme described in \cite{erik04} to prepare the initial data. With GrACE, the
time step for every mesh level is set to be that the Courant factor times the
finest spatial resolution. However, with our new infrastructure,
the time step is set by following the recipe in \cite{brugmann08,shibata08}.
Typically, the Courant factor is set between $0.25$ and $0.5$.
The results obtained with our infrastructure and with GrACE's fixed mesh
refinement are consistent with each other.

The choice of gauge condition is essential for a long-term stability in
the numerical evolution to avoid the encounter of singularity and large spatial
coordinate stretch.
The ``1+log'' condition for the lapse \cite{Bona:1994dr} and the
$\Gamma$-driver condition for the shift \cite{Alcubierre:2002kk},
which are called ``the moving puncture gauge'' together, have been successfully
applied to the long-term stable BBH simulations based on the BSSN formulation
\cite{moving_punctur1,moving_punctur2}.
In \cite{cao08}, we have reviewed and studied the parameters and the additional
advection terms for this type of gauge condition used in the community.
Therefore, in this work, we adopt
\begin{eqnarray}
\partial_t\alpha-\lambda_1\beta^j\partial_j\alpha &=& -2\alpha K,\\
\partial_t\beta^i-\lambda_2\beta^j\partial_j\beta^i &=& \frac{3}{4}f(\alpha)B^i,\\
\partial_t B^i-\lambda_3\beta^j\partial_j B^i &=&
\partial_t\tilde{\Gamma}^i-\lambda_4\beta^j\partial_j\tilde{\Gamma}^i-\eta B^i,
\end{eqnarray}
where $(f(\alpha),\eta,\lambda_1,\lambda_2,\lambda_3,\lambda_4)=
(\alpha,2,1,1,1,1)$.
The initial gauge conditions are chosen as $\alpha=\psi^{-2}$, $\beta^i=0$,
and $B^i=0$.
It has been confirmed in \cite{cao08} that this configuration allows stable
evolutions.
%%%%%%%%%%%%%%%%%%%%%%%%%%%%%%%%%%%%%%%%%%%%%%%%%%%%%%%%%%%%%%%
\subsection{Multi-puncture initial data with spectral method}\label{NPuncture}
%%%%%%%%%%%%%%%%%%%%%%%%%%%%%%%%%%%%%%%%%%%%%%%%%%%%%%%%%%%%%%%
In this section, we review the puncture scheme and then describe how the
multi-puncture BH initial data is constructed by the multi-domain spectral
method extended from the LORENE library \cite{lorene,loreneweb}. Here we
emphasize that our initial data for the three-BH problem is from the numerical
calculation solving the full Einstein constraint equations.

With the assumption of conformal flatness and maximal slicing in the conformal
decomposition of the $3+1$ formalism of general relativity
(see \cite{IDlecture} for instance), the constraint equations are greatly
simplified and also decoupled. And thus the momentum constraint allows the
Bowen-York solution \cite{BJYJ80} for the conformally trace-free part of
the extrinsic curvature of each hole,
\begin{eqnarray}
\hat{A}_a^{ij}\equiv\psi^{10}A_a^{ij}&=&\frac{3}{4r^2}[P^{(i}n^{j)}-
2(\gamma^{ij}-n^i n^j)P_k n^k] \nonumber \\
&+& \frac{3}{2r^3}n^{(i}\epsilon^{j)k\ell}S_k n_\ell,
\end{eqnarray}
%\begin{equation}
%\hat{A}_a^{ij}\equiv\psi^{10}A_a^{ij} = \frac{3}{4r^2}[P^{(i}n^{j)}-
%2(\gamma^{ij}-n^i n^j)P_k n^k] + \frac{3}{2r^3}n^{(i}\epsilon^{j)k\ell}S_k n_\ell,
%\end{equation}
where $\vec n$ is the spatial unit vector pointing away from the
puncture, and $\vec P$ and $\vec S$ correspond respectively to the
linear momentum and the intrinsic angular momentum of each hole.
$\hat{A}^{ij}$ can be linearly superposed for multi-hole spacetime with
$\hat{A}^{ij}=\sum_{a=1}^N\hat{A}_a^{ij}$ since the momentum
constraint equation is linear in this case. The conformal factor
then can be solved from the Hamiltonian constraint as
\begin{equation}
\tilde{D}^2\psi=-\frac{1}{8}\psi^{-7}\hat{A}_{ij}\hat{A}^{ij}.
\end{equation}
To deal with the physical singularity of the BH,
one separates out the singular part
\begin{equation}\label{singular}
\psi_s=1+\frac{1}{2}\sum_a\frac{m_a}{r_a},
\end{equation}
from the conformal factor $\psi$, where $m_{a}$ is the mass parameter for each
puncture and $r_{a}$ is the coordinate distance from each puncture.
Equation (\ref{singular}) is the exact multi-hole solution at rest satisfying
the flat Laplacian equation \cite{BDLR63}. The desired regular part
$u\equiv\psi-\psi_s$ satisfies the elliptic equation
\begin{equation}
\tilde{D}^2 u=-\frac{1}{8}\hat{A}^{ij}\hat{A}_{ij}(\psi_s+u)^{-7}.
\label{puncture_eq}
\end{equation}
The existence and uniqueness of the solution has been discussed in
\cite{brandt97}.

Solving elliptic equations is computationally expensive.  The spectral method
is emphatically suitable to be applied to it for its high precision and fast
convergence, provided the solution is a smooth function. The spectral method
has been successfully applied to study the relativistic star model
\cite{BSGM98}. And the LORENE library has been developed to provide a
multi-shell domain framework for this type of problem. For the puncture initial
data, due to the at most $\mathcal{C}^2$ property of $u$ at the puncture as
discussed in \cite{brandt97}, the direct application of the spectral method
would only give a polynomial convergence. The two-puncture data with spectral
method was first extensively studied in \cite{AMBT04} with a series of
coordinate transformations which modify and improve the order of smoothness of
the solution in the new coordinate. This scheme has been widely used in the
numerical relativity community, though it is not easy to be generalized to a
multi-puncture scenario. The multi-puncture data was studied recently with
a multi-grid method in \cite{GPBC10}.
%%%%%%%%%%%%%%%%%%%%%%%%%%%%%%%%%%%%%%%%%%%%%%%%%%%%%%%%%%%%%%%
\begin{figure}[tbp]
\centering\includegraphics[width=\columnwidth]{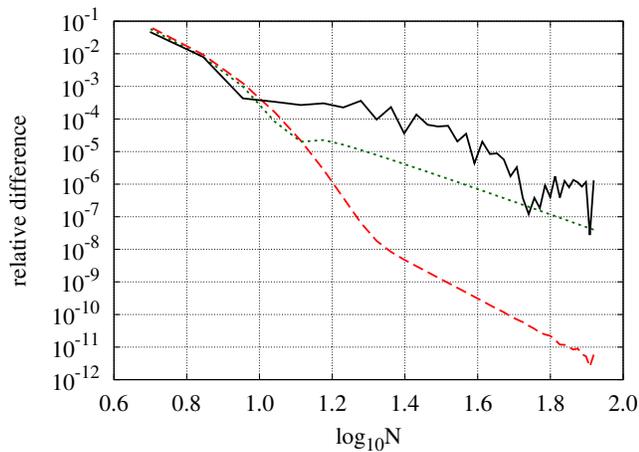}
\caption{Local convergence test (at the puncture) for the initial data of
a single boosted BH (dashed-red) with $P=0.2$, a spinning BH (dotted green)
with $S=0.2$, and an equal-mass BBH (solid) separated by $6M$ with $P=\pm0.2$,
respectively.
The vertical axis indicates the relative difference between the solution $u$ at
$N_r=N$ and the one at the next higher resolution.
For the single BH cases, the angular resolution $(N_\theta,N_\phi)=(N_r,4)$
which provides sufficient polar angular resolution for this axisymmetric test.
For the BBH case, $(N_\theta,N_\phi)=(13,16)$.
Both of them displays exponential convergence at lower resolution and
polynomial convergence after the difference has dropped below certain value
which is dependent on the choice of domain boundaries.}
\label{fig:IDconvergence}
\end{figure}
%%%%%%%%%%%%%%%%%%%%%%%%%%%%%%%%%%%%%%%%%%%%%%%%%%%%%%%%%%%%%%%

Our multi-puncture initial data solver is motivated by the earlier work
\cite{GEGB02} for the excised BBH initial data. In this code we cover on each
BH a spherical multi-shell domain, and split $u=\sum_a u_a$ (the index $a$ runs
over the number of the BHs) and the puncture equation (\ref{puncture_eq}) into
\begin{equation}
\tilde{D}^2 u_a=-\frac{1}{8}\hat{A}_a^{ij}\hat{A}_{ij}(\psi_s+u)^{-7}.
\end{equation}
Note that only one of the extrinsic curvature tensors is split
in the equation. Therefore, the source term in the RHS contributes
only near each hole. The use of spherical polar coordinates is
adequate for solving the equation near the punctures. Both $\hat{A}_{ij}$
and $\psi_s$ are known analytically and would be set once and for all
in the source term. The value of $u_a$ is imposed to be zero at the physical
outer boundary (at infinity) on the outermost, compactified shell.
And $\partial_r u_a$($r=0$)$=0$ is also ensured at the punctures as the
inner boundary condition. These equations for each hole are then
solved iteratively with the Poisson solver in the LORENE library
until each successive difference of $\delta u_a$ is as small as
possible, typically $10^{-11}$. In the three-BH scenario,
the massive hole is fixed and offers a fixed gravitational potential
background, therefore there is only two Poisson
equations that need to be solved in each iteration. We have also
checked the resulting data and found that it is very close to the one from the
superposition of the solution of the small binary and the solution of the
third massive BH. This is expected since the massive BH in our study is
distant to the binary.

The convergence tests shown in Fig.~\ref{fig:IDconvergence} for a single
boosted BH, a spinning BH, and a BBH display a rapid convergence to a high
precision.
It is clear that the convergence is exponential with low resolution,
but turns to be polynomial with high resolution, as expected in \cite{jpbo01}.
The turning point from an exponential convergence to a polynomial
convergence mainly depends on the choice of the domain boundaries,
as well as the order of the solution's smoothness.
However, it is also known that the polynomial convergence at higher resolution
depends only on the singular structure of the solution.

The momentum parameter for the quasi-circular binary is set according to the
fitting curve based on the helical Killing vector conditions in
\cite{tichy04}. As we note that, in the puncture scheme, only the metric of
3-geometry was specified, while the initial lapse and shift are chosen as the
moving puncture gauge condition described in the beginning of this section.
Nevertheless, it is justified in our experience that the choice of initial
gauge for the puncture data has little effect on the resulting physical
content, at least within our numerical accuracy.
%%%%%%%%%%%%%%%%%%%%%%%%%%%%%%%%%%%%%%%%%%%%%%%%%%%%%%%%%%%%%%%
\subsection{Boundary condition} \label{secbc}
%%%%%%%%%%%%%%%%%%%%%%%%%%%%%%%%%%%%%%%%%%%%%%%%%%%%%%%%%%%%%%%
In numerical relativity the Sommerfeld radiation boundary condition
is widely used and also a good approximation for no reflection
from the boundary at a finite distance. For our BBH simulation with
a time-independent gravitational background, the field variable set
$Q(t,r)=\{\phi,\tilde\gamma_{ij}-\eta_{ij},
K,\tilde{A}_{ij},\tilde\Gamma^i\}$ receives the
gravitational contribution from the BBH and from the third large BH
respectively as
\begin{equation}
Q(t,r)=Q_{\rm BBH}(t,r)+Q_{\rm 3rd}(r),
\end{equation}
where the first term $Q_{\rm BBH}$ in the RHS includes
the dynamical outgoing wave, while the second term $Q_{\rm 3rd}$
in the RHS serves as a fixed gravitational background which are trivially zero
except the conformal factor of the third BH,
\begin{equation}
\phi_{\rm 3rd}=\ln\left(1+\frac{m_3}{2|\vec r-\vec{r}_3|}\right).
\end{equation}
In order to update the variable set $Q(t,r)$ at the outer boundary,
we first extract $Q_{\rm BBH}(t_0,r)$ at the current time step $t_0$
by subtracting from $Q(t_0,r)$ the $Q_{\rm 3rd}(r)$ which is fixed
in the whole evolution. Then, as usual, the Sommerfeld boundary
condition is applied \cite{shibata_naka95} to obtain the numerical data of $Q_{\rm
BBH}(t_1,r)$ on the outer boundary for the next time step $t_1$. And
finally we add $Q_{\rm 3rd}(r)$ back to the new $Q_{\rm BBH}(t_1,r)$
to obtain $Q(t_1,r)$.
For the gauge variables, we use the usual Sommerfeld boundary condition for 
stability.

To justify the appropriateness of the modified boundary condition,
we have performed two simulations with different outer boundaries and found
their difference on the trajectory and gravitational waveform of the BBH are
ignorable, as shown in Fig.~\ref{fig13} (Further details are explained in
Sec.~\ref{inor}). Therefore, our boundary condition does not introduce
unphysical drift to the BBH and the effect observed in this work is indeed
resulted from the third BH.
%%%%%%%%%%%%%%%%%%%%%%%%%%%%%%%%%%%%%%%%%%%%%%%%%%%%%%%%%%
\section{Numerical results}\label{seciii}
%%%%%%%%%%%%%%%%%%%%%%%%%%%%%%%%%%%%%%%%%%%%%%%%%%%%%%%%%%
\begin{figure}[tbp]
\centering\includegraphics[width=\columnwidth]{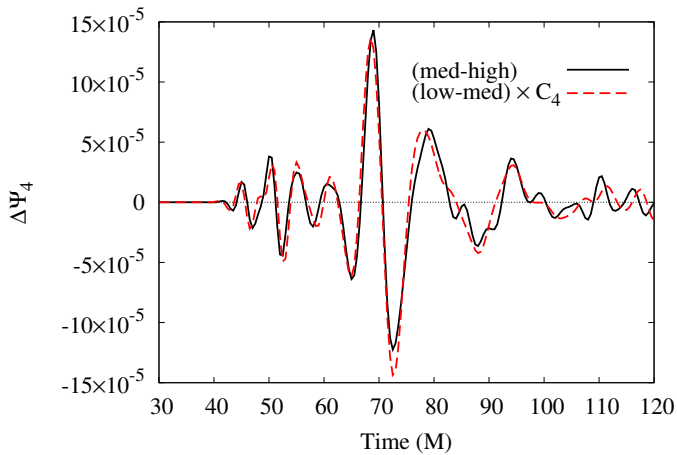}
\caption{Fourth-order convergence of the waveforms for the head-on
collisions of an isolated BBH system. The solid line represents the
difference of $\Psi_4$ between the high and medium resolutions. The
red-dashed line represents the difference between the medium and low
resolutions with a convergence factor $C_4$.} \label{ho_convergence}
\end{figure}
%%%%%%%%%%%%%%%%%%%%%%%%%%%%%%%%%%%%%%%%%%%%%%%%%%%%%%%%%%
There is an obvious hierarchical structure in the considered system,
as shown in Fig.~\ref{fig1}. A small BBH system and a third BH,
which is distant and much more massive, form a bigger binary system.
From the point view of eccentricity $e$ of a binary, there are two
limits of the possible configurations existing in the universe : a
head-on collision ($e=1$) and (quasi-)circular orbit ($e=0$). Here
we would like to investigate four possible combinations of the
limiting cases in order to offer some clues for understanding this
kind of astrophysical system.

The considered hierarchical structure comprises four possible cases.
They are (A) the head-on-freefall case: a BBH is in the process of a head-on
collision while its center of mass (CoM) freefalls toward a third BH;
(B) the head-on-orbiting case: a BBH is in the process of a head-on collision
while its CoM circularly orbits a third BH;
(C) the inspiral-freefall case: a BBH is in the process of a quasi-circular
inspiral while its CoM freefalls toward a third BH;
(D) the inspiral-orbiting case: a BBH is in the process of a quasi-circular
inspiral while its CoM circularly orbits around a third BH.
The numerical results of the four scenarios will be presented in the
following subsections. The total mass of the BBH in these cases is
set to be $M=1$.
%%%%%%%%%%%%%%%%%%%%%%%%%%%%%%%%%%%%%%%%%%%%%%%%%%%%%%%%%%
\begin{figure}[tbp]
\centering\includegraphics[width=\columnwidth]{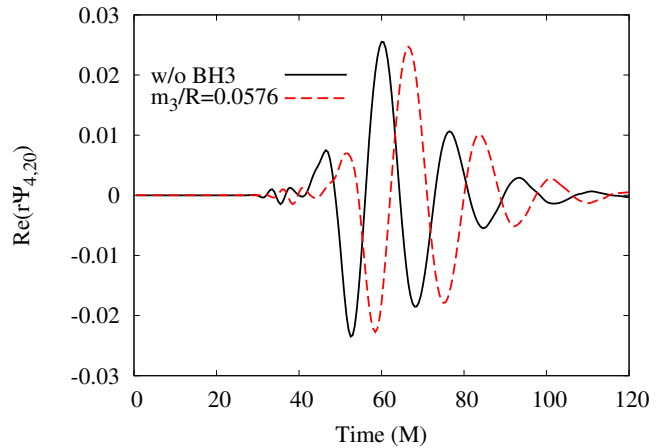}
\caption{The typical waveforms for the head-on collisions of an
isolated BBH and for the head-on-freefall case in Sec.~\ref{hoho}.
The red-dashed (black-solid) line represents the case with (without)
the third BH. The $(\ell=2, m=0)$ mode of $\Psi_4$ detected at $r=40$ is
plotted. The gravitational potential of the third BH
causes mainly shifts in time, decreases in amplitude, and
prolongation of the wavelength in the gravitational waveform.}
\label{fig2}
\end{figure}
%%%%%%%%%%%%%%%%%%%%%%%%%%%%%%%%%%%%%%%%%%%%%%%%%%%%%%%%%%%%%%%

We use the similar grid setup as in our previous work \cite{cao08}.
There are five levels of grid for the fixed mesh refinement.
The physical boundary is put at $64M$. The domain takes a cubic shape,
and the length of the domain in each level shrinks to one half of the length of
the preceding level.
For the finest level, the domain contains two boxes: they are bounded by
($-2<x<2$, $-2<y<2$, $0.25<z<4.25$) and ($-2<x<2$, $-2<y<2$, $-4.25<z<-0.25$),
and are movable with the moving-box technique.
Before the study on the effect of a third BH, the convergence of our code for
an isolated BBH in a head-on collision is tested.
The fourth order convergence of $\Psi_4$ is shown in Fig.~\ref{ho_convergence}
with the convergence factor
\begin{equation}
C_4=\frac{(h_1/h_2)^4-1}{1-(h_3/h_2)^4}\approx 0.65,
\end{equation}
where $h_1=1/32$, $h_2=1/28$ and $h_3=1/24$ are the finest grid widths for the
high, median and low resolution, respectively. Some minor disagreement for
$t>80$ was expected to come from the outer boundary which could give a
second-order error.

In the study of the BBH's evolution under a gravitational background,
we use the Newman-Penrose scalar $\Psi_4$ instead of the gravitational strain
$h^{+/\times}$ to represent the gravitational wave. $\Psi_4$ is widely used
in numerical relativity community as a measurement of gravitational waveform.
So it is convenient to use $\Psi_4$ and its spherically harmonic decompositions,
instead of $h^{+/\times}$, for comparison with previous works.
Furthermore, $\Psi_4$ can be linked to $h^{+/\times}$ straightforwardly as
explained in \cite{frans07,RANT08}.
%%%%%%%%%%%%%%%%%%%%%%%%%%%%%%%%%%%%%%%%%%%%%%%%%%%%%%%%%%%%%%%
\subsection{Head-on-freefall case}\label{hoho}
%%%%%%%%%%%%%%%%%%%%%%%%%%%%%%%%%%%%%%%%%%%%%%%%%%%%%%%%%%%%%%%
We will consider, in this and the next subsections, the BBHs in head-on
collisions while their CoMs either free-fall toward or circularly orbit
around the third massive BH. The massive BH has the mass $m_3$ and is located
at $(-R,0,0)$. We will vary the mass $m_3$ while fix the coordinate distance at
$R=1000$, and the ratio $m_3/R$ will represent the strength of different
gravitational potential due to the third massive BH \cite{CITE1}.

%%%%%%%%%%%%%%%%%%%%%%%%%%%%%%%%%%%%%%%%%%%%%%%%%%%%%%%%%%%%%%%
\begin{figure}[tbp]
\centering\includegraphics[width=\columnwidth]{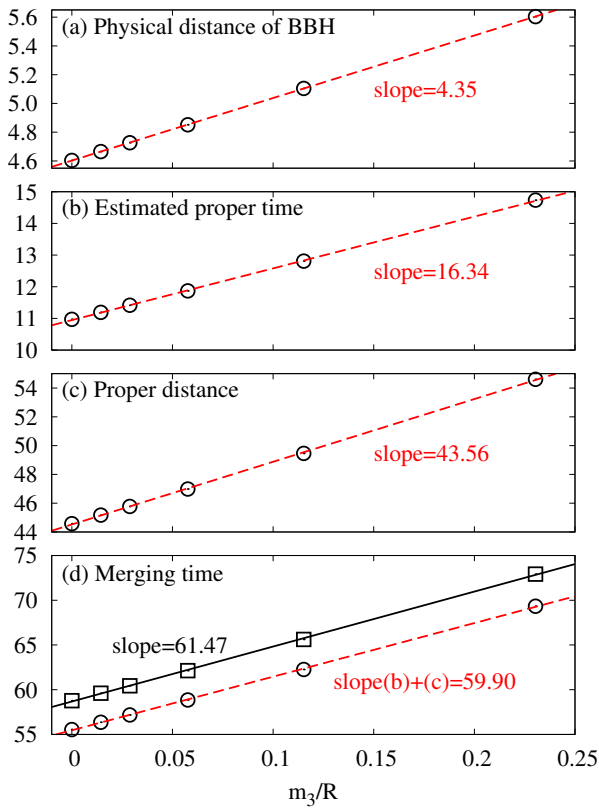}
\caption{(a)Initial physical distance between the small BBH,
(b)proper merging time from the initial physical distance as
estimated with Eq.~(\ref{newton}), and (c)proper distance from the center
of the small BBH system to the detector, respectively, with respect
to the strength of the gravitational potential of the third BH. (d)
Comparison of the proper merging time, which are either measured at
the detector(squares) or as the direct addition(circles) of (b) the
(c). The circles and squares indicate the numerical data. The slopes
of their linear fitting are also given.} \label{fig3}
\end{figure}
%%%%%%%%%%%%%%%%%%%%%%%%%%%%%%%%%%%%%%%%%%%%%%%%%%%%%%%%%%%%%%%
In Fig.~\ref{fig2}, we compare the gravitational waves from the head-on
collision BBH affected by a third BH and from an isolated binary in our
earlier result \cite{cao08}. 
The BBH in both scenarios has the mass parameters $m_1=m_2=0.5$ and
is initially located at $(0,0,\pm 1.1515)$, which are about the ISCO.
We can see from this figure that the gravitational potential of the
third BH causes a time delay, a slight decrease of the amplitude
of the $\ell=2$ mode of $\Psi_4$
and a prolongation of the wavelength in the waveform.
We also observe at the same time the excitation of higher order modes with $\ell>2$.
The decrease in the amplitude mainly comes from the gravitational redshift
caused by the third BH. Similar to the nonlinear redshift effects of an electromagnetic wave
in a medium, the potential of the third BH broadens the spectrum of
the gravitational wave instead of simple linear spectral redshift.

%\sout{
%The decrease in the amplitude mainly comes from the gravitational redshift
%caused by the third BH, which not only broadens the spectrum of the
%gravitational wave but also causes the spectral redshift, 
%as the extra gravitational potential changes the background spacetime from
%one medium (without gravitational background) into another (with gravitational
%background) for the propagation of the gravitational wave. 
%}
%\sout{
%The effect of spectral-broadening is observed in our cases to decrease the amplitude
%of the $\ell > 2$ mode of $\Psi_4$ , and at the same time to excite the $\ell > 2$ modes.
%}
%\sout{The spectral-broadening phenomenon is
%similar with the one from the frequency dispersion for a
%non-monochromatic wave in electrodynamics {Jackson99}}. 
The time delay comes from three factors: (i) the delayed
merger; (ii) the prolonged proper distance from the source to the
detector and (iii) the change in the coordinate time. It can be
understood that the last one is simply caused by the change of time
gauge due to the existence of the third BH, therefore we only
discuss the first two factors in detail.
%%%%%%%%%%%%%%%%%%%%%%%%%%%%%%%%%%%%%%%%%%%%%%%%%%%%%%%%%%%%%%%
\begin{figure*}[tbp]
\begin{tabular}{rl}
\includegraphics[width=0.49\textwidth]{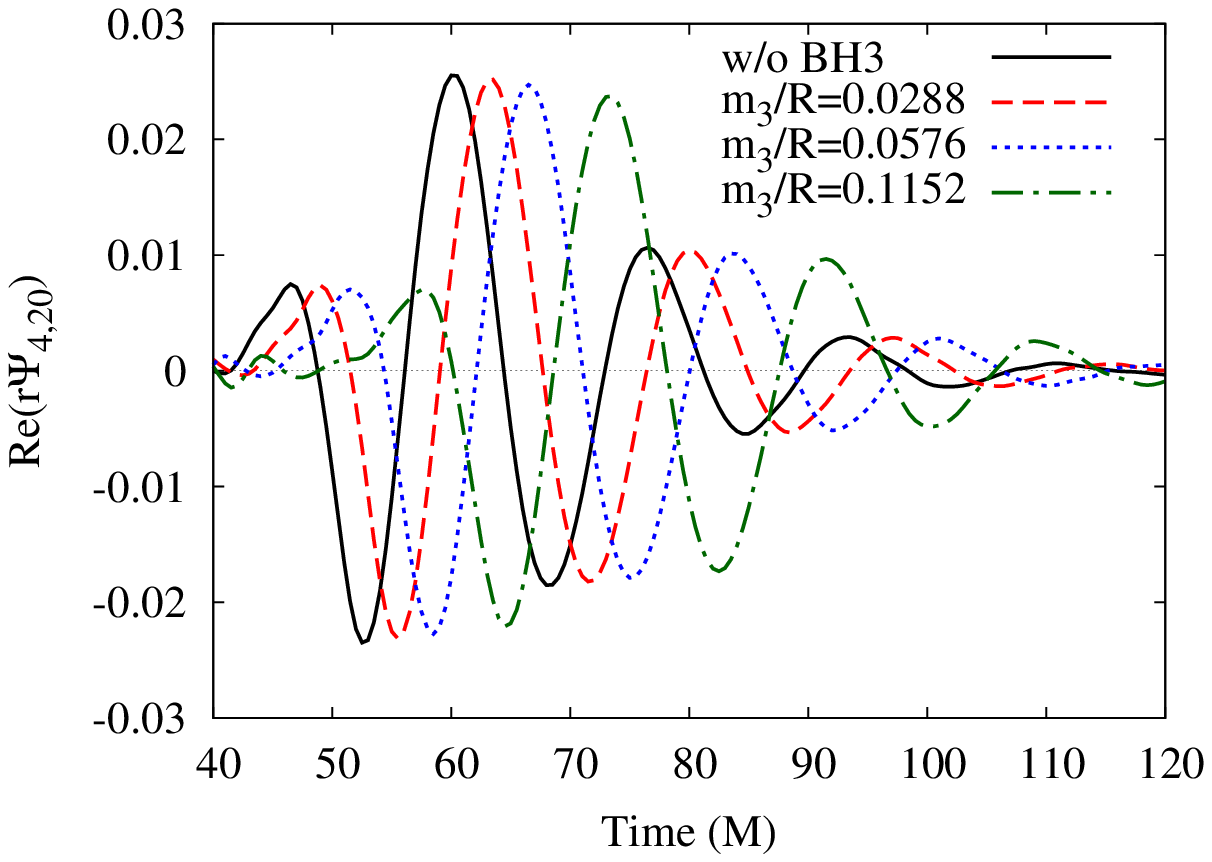} &
\includegraphics[width=0.49\textwidth]{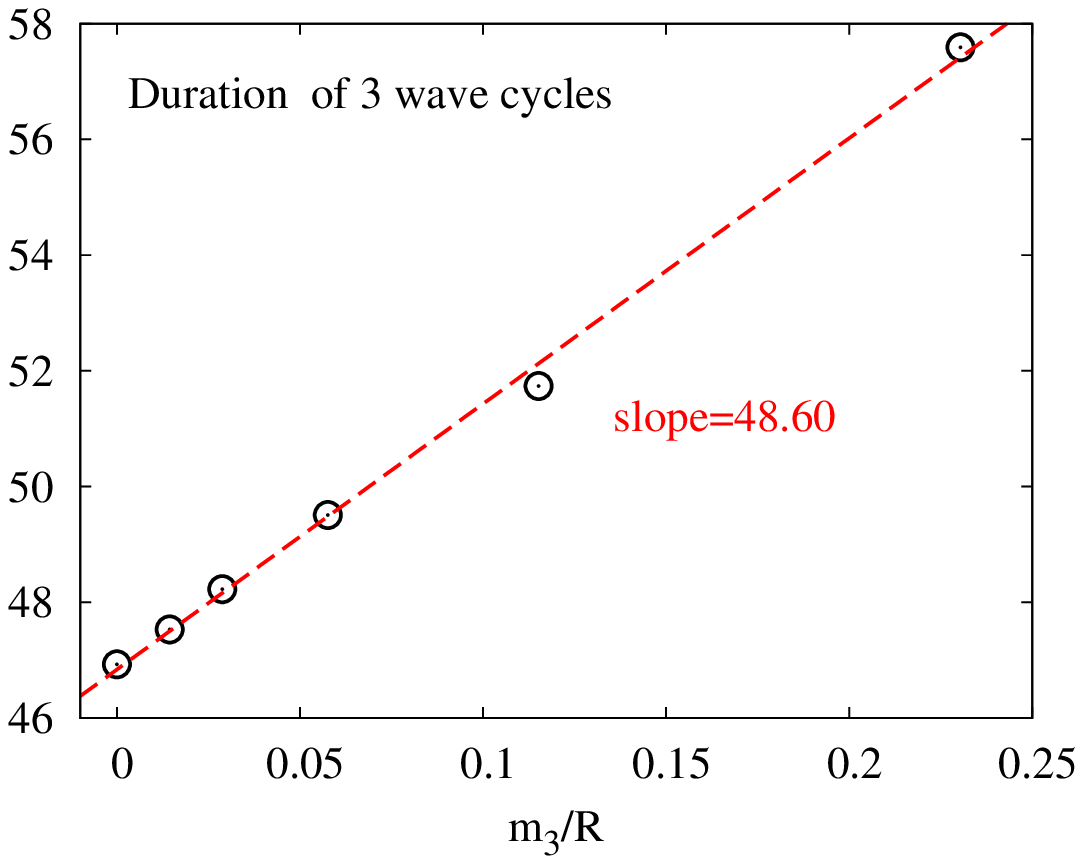}
\end{tabular}
\caption{Left panel: Waveforms of the head-on-freefall case
in Sec.~\ref{hoho} for different potential strengths of the third BH.
The amplitude decreases as the third BH's gravitational potential increases.
Right panel: duration for three wavelengths starting from the wave
peak at around $t=45$ to the peak at around $t=90$.
The wavelength increases as the third BH's gravitational potential increases.}
\label{fig4}
\end{figure*}
%%%%%%%%%%%%%%%%%%%%%%%%%%%%%%%%%%%%%%%%%%%%%%%%%%%%%%%%%%%%%%%
\begin{figure}[tbp]
\centering\includegraphics[width=\columnwidth]{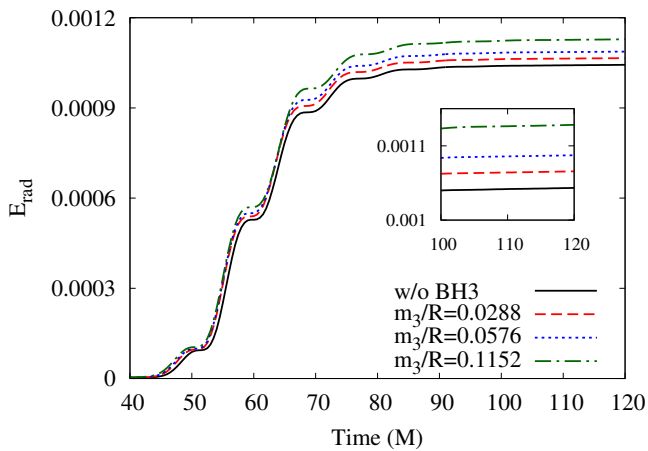}
\caption{Accumulated radiated energy with respect to time for different
gravitational potential strengths.
A stronger gravitational potential results in more energy radiation.}
\label{fig5}
\end{figure}
%%%%%%%%%%%%%%%%%%%%%%%%%%%%%%%%%%%%%%%%%%%%%%%%%%%%%%%%%%%%%%%

Factor (i) comes from the effectively larger proper separation, i.e.,
the physical distance $d$ of the BBH, due to the curved spacetime background
of the third BH. Here the physical distance $d$ is defined as the proper length
of the shortest line connecting two individual apparent horizons, which is
along the $z$-axis in the current case, and is explicitly calculable at the
initial slice with $d=\int\sqrt{\gamma_{ij}{\rm d}x^i{\rm d}x^j}$ provided
these two
apparent horizons are found. We plot $d$ in first panel of Fig.~\ref{fig3} with
respect to the potential strength of the third BH. Unlike the coordinate
merging time of the BBH that can be obtained from the simulation, it is
ambiguous to evaluate the proper merging time because of the singularities in
the BBH. Due to our perturbational treatment of the third BH's gravitational
effect and the small velocity (compared with speed of light) of each BH in the
head-on BBH, Newtonian mechanics can provide a rough estimate of the
proper merging time for a given $d$ by virtue of the ordinary equation,
\begin{equation}\label{newton}
\frac{{\rm d}^2}{{\rm d}t^2}\frac{\vec d}{2}=-m\frac{\vec d}{d^3}
-m_3\frac{\vec{\mathcal D}+\vec d/2}{|\vec{\mathcal D}+\vec d/2|^3}\approx
-m\frac{\vec d}{d^3},
\end{equation}
where $m$ is the mass of individual BH in the BBH, and the physical distance
between the small binary and the third BH is $\mathcal D$. In the last term
of the above equation, $\mathcal D$ does not appear in the final expression.
However, the third BH's
gravity still affects the physical distance $d$ through its effect on the 
3-metric. We integrate the above
equation to obtain the estimated proper merging time for certain $d$, and show
the results in Fig.~\ref{fig3}b. For the factor (ii), the proper distance
between the BBH system and the detector is prolonged due to the potential of
the third BH. In Fig.~\ref{fig3}c, the prolonged proper distance
from the BBH to the detector was plotted with respect to the
potential strength of the third BH. The prolonged proper distance should be
equal to the proper time as the gravitational wave propagates in the speed of
light. The two effects responsible for the proper time delay, one in
the merger of head-on collision and the other in the propagation of the
gravitational wave, can account for the numerically detected time delay
for the gravitational wave. To verify this point, we measure the proper
time of the highest peak of the gravitational wave arriving the
detector(squares in Fig.~\ref{fig3}d), and compare it with the result from
the direct addition of the estimated proper merging time and the
proper propagation time(circles in Fig.~\ref{fig3}d) as mentioned.
After a linear fit to get the slope as the time delay rates with
respect to the extra potential, it turns out that the measured
proper time delay rate, $61.47$ is very close to the sum of the
delay rate of the estimated proper merging time and the proper
propagation time, as shown in the Fig.~\ref{fig3}d \cite{CITE2}.

As mentioned earlier, the redshift effect due to the gravitational potential
of the third BH will decrease the amplitude of $\Psi_4$'s $\ell=2$ mode as well
as prolonging the wavelength of the gravitational waveform.
They are clearly shown in Fig.~\ref{fig4}, where the stronger gravitational
potential causes smaller amplitude as in the left panel, and the longer
durations of three cycles (from the first to the third peak in the left panel),
i.e., the longer wavelength as shown in the right panel.

%%%%%%%%%%%%%%%%%%%%%%%%%%%%%%%%%%%%%%%%%%%%%%%%%%%%%%%%%%%%%%%
\begin{figure*}[tbp]
\begin{tabular}{rl}
\includegraphics[width=0.49\textwidth]{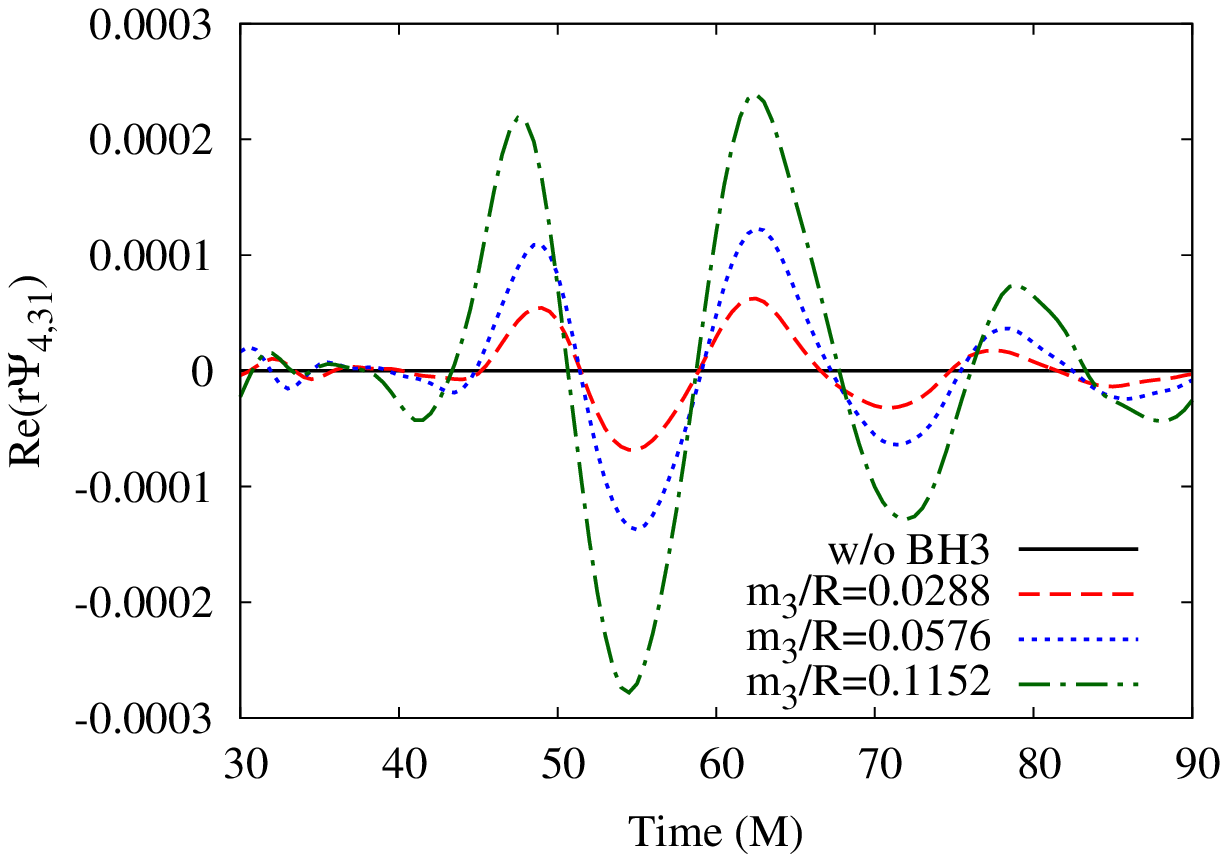} &
\includegraphics[width=0.49\textwidth]{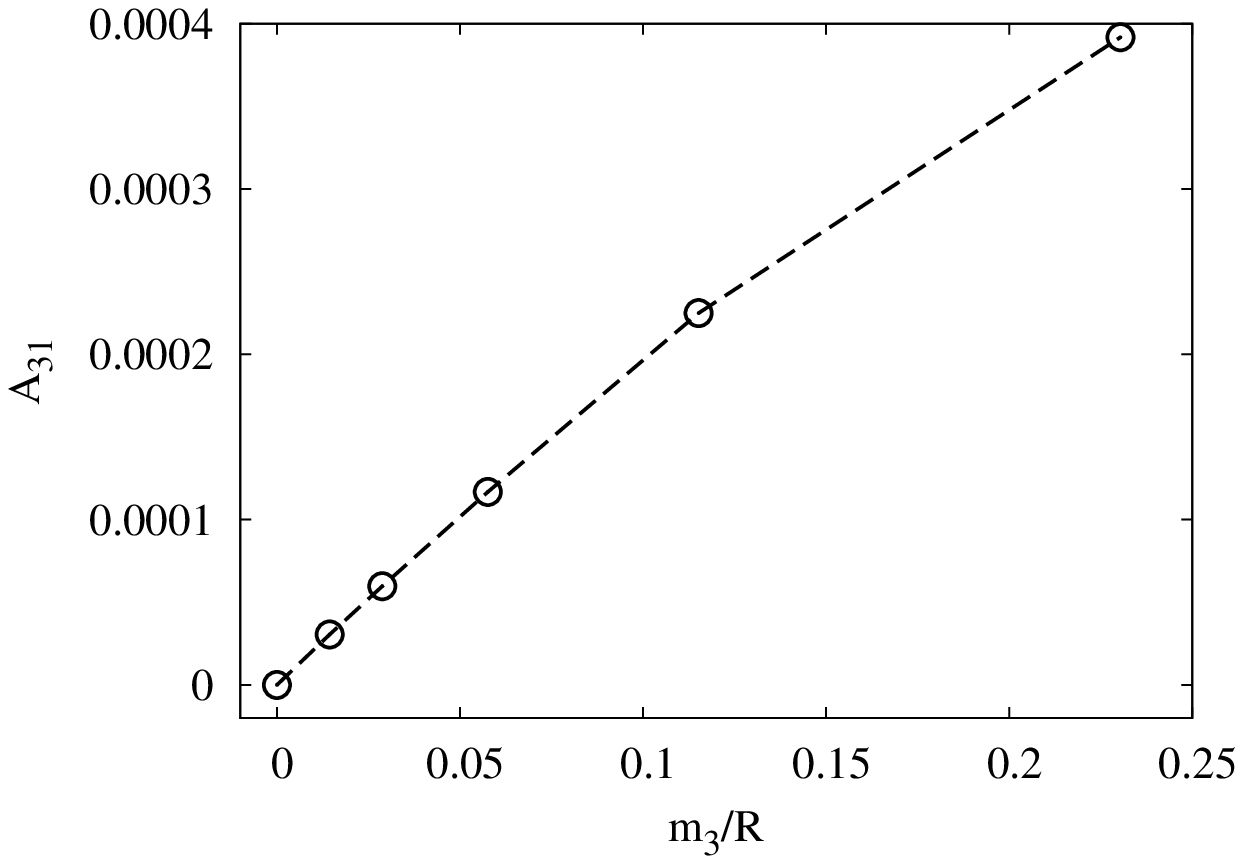}
\end{tabular}
\caption{Left panel: Waveforms of the
($\ell=3$, $m=1$) mode of $\Psi_4$ for different background potential strengths.
Right panel: The largest amplitude of $\Psi_{4,31}$ showing a nonlinear growth pattern.}
\label{fig6}
\end{figure*}
%%%%%%%%%%%%%%%%%%%%%%%%%%%%%%%%%%%%%%%%%%%%%%%%%%%%%%%%%%%%%%%
\begin{figure}[tbp]
\centering\includegraphics[width=\columnwidth]{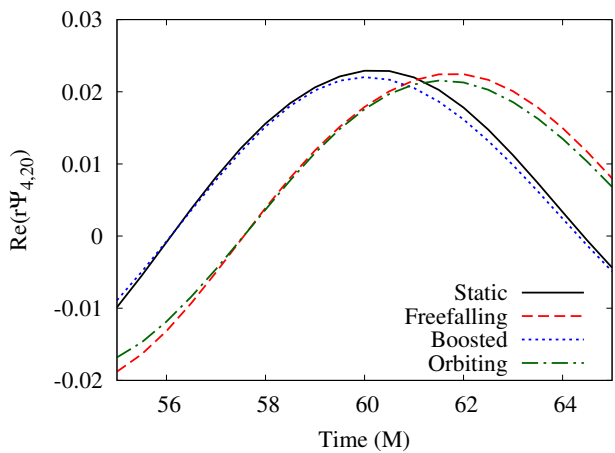}
\caption{Waveforms of ${\rm Re}(r\Psi_{4,20})$ for four scenarios mentioned in
Sec.~\ref{ho_orbiting}. These results show that the relativistic Doppler
effect on the behavior of the gravitational waveform is much less important
than the gravitational redshift effect in the head-on-orbiting case.}
\label{fig7}
\end{figure}
%%%%%%%%%%%%%%%%%%%%%%%%%%%%%%%%%%%%%%%%%%%%%%%%%%%%%%%%%%%%%%%
\begin{figure*}[tbp]
\begin{tabular}{rl}
\includegraphics[width=0.49\textwidth]{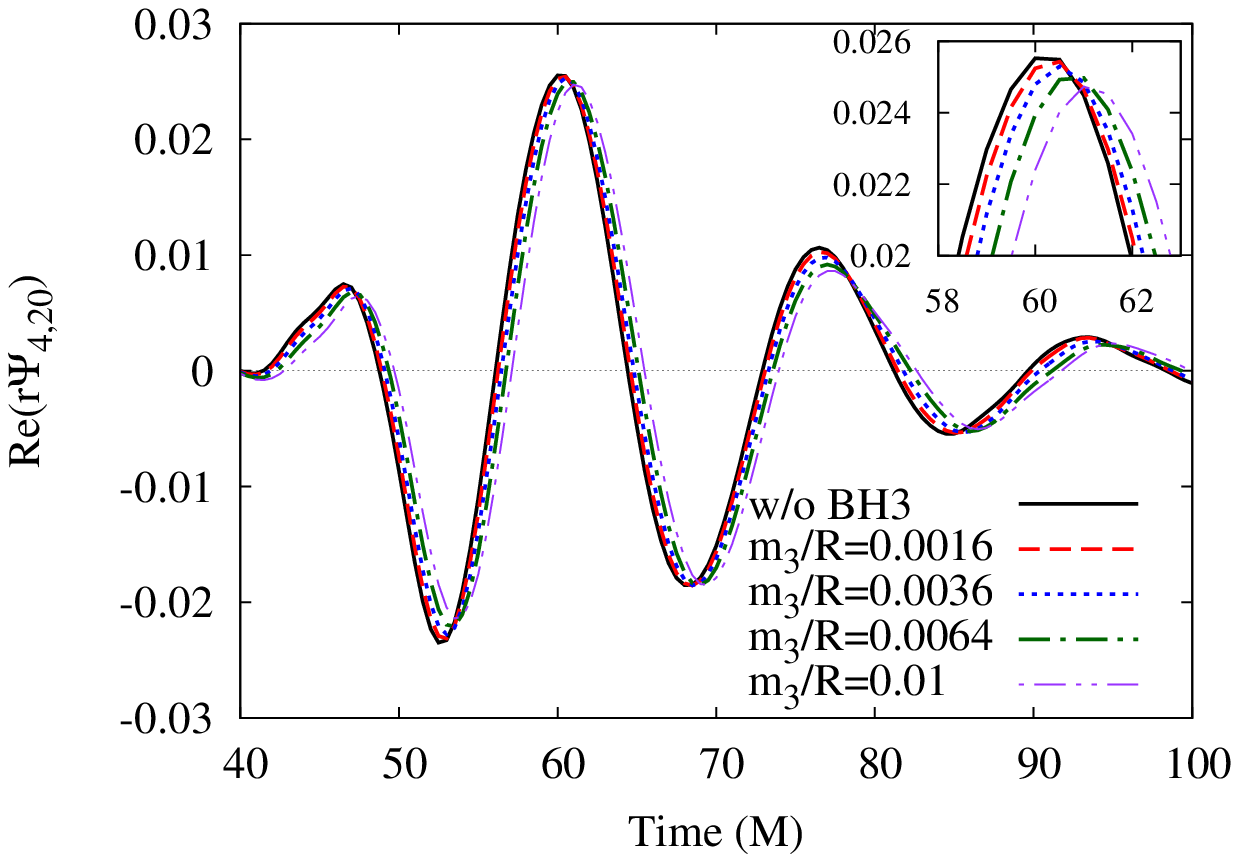} &
\includegraphics[width=0.49\textwidth]{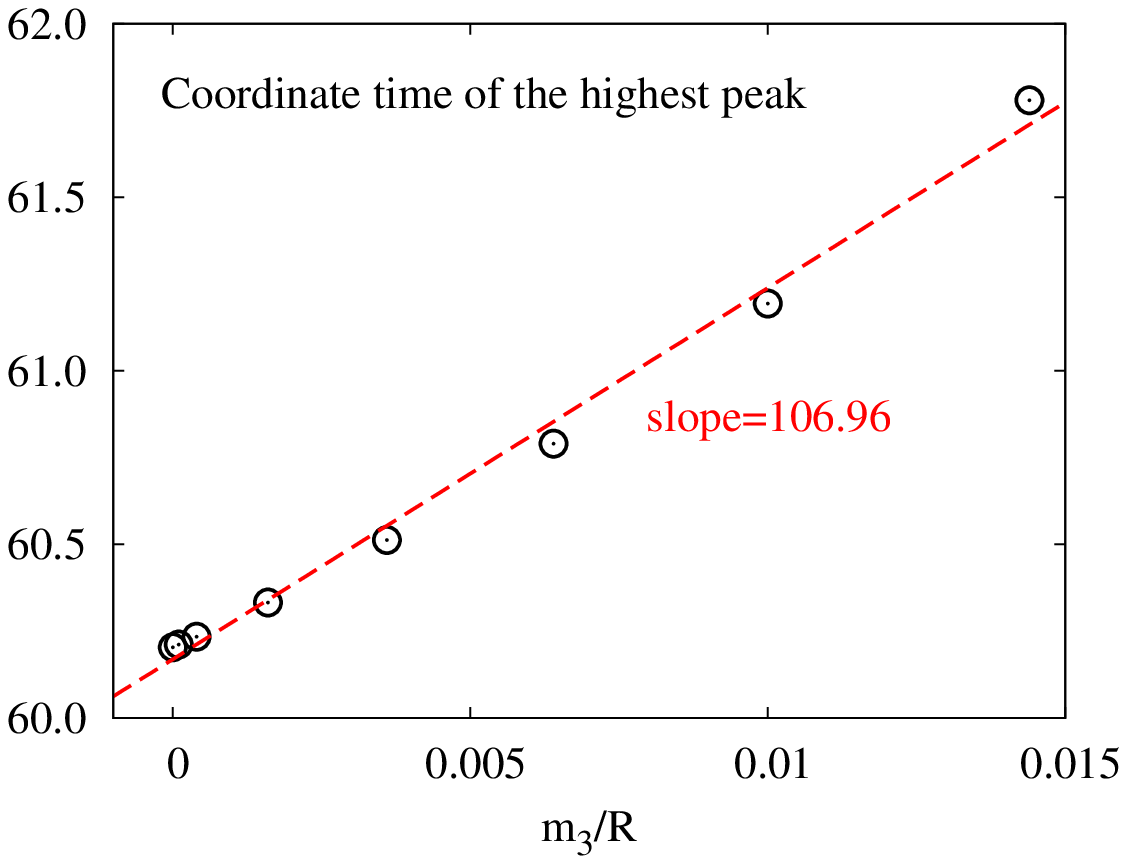}
\end{tabular}
\caption{Left panel: Waveforms of the head-on-orbiting case in
Sec.~\ref{ho_orbiting} for different background potential strengths.
They are qualitatively similar to Fig.~\ref{fig4}, except the smaller
decreasing amplitude due to the smaller background potential strength.
Right panel: coordinate time of the highest peak of $\Psi_{4,20}$ arriving the
detector for different background potential strengths.}
\label{fig8}
\end{figure*}
%%%%%%%%%%%%%%%%%%%%%%%%%%%%%%%%%%%%%%%%%%%%%%%%%%%%%%%%%%%%%%%
We also investigate the gravitational energy radiation for this case and show
the result in Fig.~\ref{fig5}. It shows that a stronger gravitational potential
results in more energy radiation.
This can be understood as follows: when the small BBH system is located in
the gravitational potential of a third BH,
the potential energy will be put into the binary system due to the interaction,
and some part of this energy is carried out by the gravitational wave.
Then the radiation energy during the head-on collision is enhanced by the
potential of the third BH.
Campanelli {\it et al.} \cite{campanelli06} have studied
the ISCO problem for triple BH systems. They concluded that the
third BH will enhance the gravitational radiation of BBH,
which is consistent with our numerical result. In our perturbational
treatment, the curved background is fixed and thus there is no
back-reaction between the BBH and the third BH. In a more realistic
case, the back-reaction will speed up the free-falling process of
the BBH toward the third BH.

Besides distorting the $\ell=2$ modes,
the existence of the third BH also induces higher-order modes,
especially the $\ell=3$ modes, in the BBH's gravitational radiation.
The left panel of Fig.~\ref{fig6} shows the waveform of the
$(\ell=3$, $m=1)$ mode of the gravitational radiation induced in the current
case with respect to the different strength of the gravitational potential.
The right panel of Fig.~\ref{fig6} shows the amplitudes of
the $(\ell=3$, $m=1)$ mode of the gravitational wave,
which demonstrates a nonlinear growth with respect to the strength of the third
BH's gravitational potential.
In principle, it is possible for the nonlinear growth of the $\ell=3$
mode to be used as a signature to distinguish a three-BH system from a BBH
system.
However, the phenomenon could also complicate the identification of the source
of the gravitational wave.
Even in the Newtonian framework, certain three-body system with the same
quadrupole wave forms may have quite distinguishable higher-order modes of the
gravitational wave \cite{torigoe09}.
Without a careful understanding on the waveform pattern,
the gravitational wave emitted from a three-BH system,
like the cases described in this work,
could be misidentified as one from an unequal-mass BBH system in which the
higher-order modes of gravitational wave also exist.
%%%%%%%%%%%%%%%%%%%%%%%%%%%%%%%%%%%%%%%%%%%%%%%%%%%%%%%%%%%%%%%
\subsection{Head-on-orbiting case}\label{ho_orbiting}
%%%%%%%%%%%%%%%%%%%%%%%%%%%%%%%%%%%%%%%%%%%%%%%%%%%%%%%%%%%%%%%
In this subsection, we study the head-on collision of a BBH while the BBH
system moves along a circular orbit around a third large BH.
The orbiting velocity of
the center of mass (CoM) of the BBH is initially set according to the 
Newtonian gravity, i.e., $v^2\approx m_3/R$. This treatment is valid
since the third massive BH is distant to the BBH and the orbit of
the BBH is set to be circular. In fact, the massive BH are so
distant such that by the end of merger of the BBH, the CoM
displacement of BBH is only a little fraction of one big orbit,
which makes the eccentricity of the big orbit not relevant to our
study. Given the linear momentum parameters and the position
parameters of the BHs, we solve the puncture initial data as
described in Sec.~\ref{NPuncture}. For the third large BH, we fix
the coordinate distance at $R=1000$ as in the previous subsection,
and change the gravitational potential $m_3/R$ by varying the mass
parameter $m_3$.

There are two effects for a BBH orbiting in the gravitational
potential well of a third large BH: the relativistic Doppler effect
and the gravitational redshift. In order to identify the importance
of these two effects on the time delay in the gravitational waveform
in the current case, we compare the waveforms obtained from the
following four scenarios: (i) the head-on collision of an isolated
BBH; (ii) the head-on collision of a BBH freefalling towards a third
large BH; (iii) the head-on collision of a boosted isolated BBH;
(iv) the head-on collision of a BBH orbiting around a third large
BH. Here the mass parameter of the third BH is set to be $m_3=14.4$,
the same in the cases of (ii) and (iv), and the boosting
velocity in (iii) is set to be the same as the orbiting velocity in
(iv), the magnitude of which is $0.12$, according to $v^2\approx
m_3/R$ and $R=1000$. The results in Fig.~\ref{fig7} show that the
waveforms of cases (i) and (iii) are close to each other. So are the
waveforms of cases (ii) and (iv). This indicates that the gravitational
redshift of the third large BH should dominate the effect on the
time delay of the gravitational wave in the merging process of a
BBH. On the contrary, the relativistic Doppler effect has only minor
influence on the time delay of the waveform. Therefore we will
neglect in the following discussion the relativistic Doppler effect
due to the small orbiting velocity of the CoM of a BBH around a
third large BH, and focus mainly on its gravitational redshift effect.

In order to let the BBH move slow enough such that it will stay
inside the computational domain during the merger process, and thus we can
keep the necessary accuracy during the generation of gravitational wave, the
gravitational potential strength of the third BH is set to be
smaller than in the previous subsection. The numerical results for
this case show that all phenomena including the time delay of
the waveform, the decrease of the wave amplitude, the prolongation
of the wavelength, the enhancement of the energy radiated, and the
excitation of the higher-order modes are qualitatively similar to
the ones in the previous case. Therefore we only show here the
decrease in amplitude of the gravitational waveform and the
coordinate time delay of the highest peak of $\Psi_{4,20}$ with
respect to different potential strengths in Fig.~\ref{fig8}. It
can be seen in the left panel that the decrease in amplitude of the
waveform is less than that obtained in the previous subsection due
to the smaller gravitational potential strength of the third BH.
Just as in the previous subsection, the orientation of the small BBH
barely affects the result in this case because of the negligible tidal effect
from the third BH in a head-on collision.
%%%%%%%%%%%%%%%%%%%%%%%%%%%%%%%%%%%%%%%%%%%%%%%%%%%%%%%%%%%%%%%
\subsection{Inspiraling-freefall case}\label{infreefall}
%%%%%%%%%%%%%%%%%%%%%%%%%%%%%%%%%%%%%%%%%%%%%%%%%%%%%%%%%%%%%%%
\begin{figure*}[tbp]
\begin{tabular}{rl}
\includegraphics[width=0.49\textwidth]{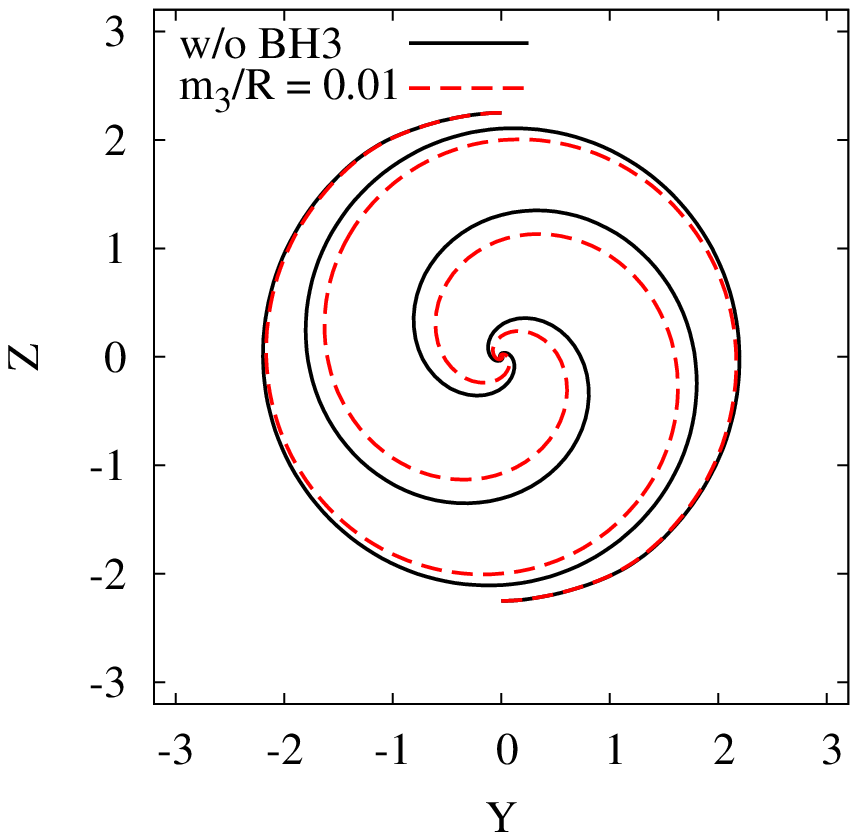} &
\includegraphics[width=0.49\textwidth]{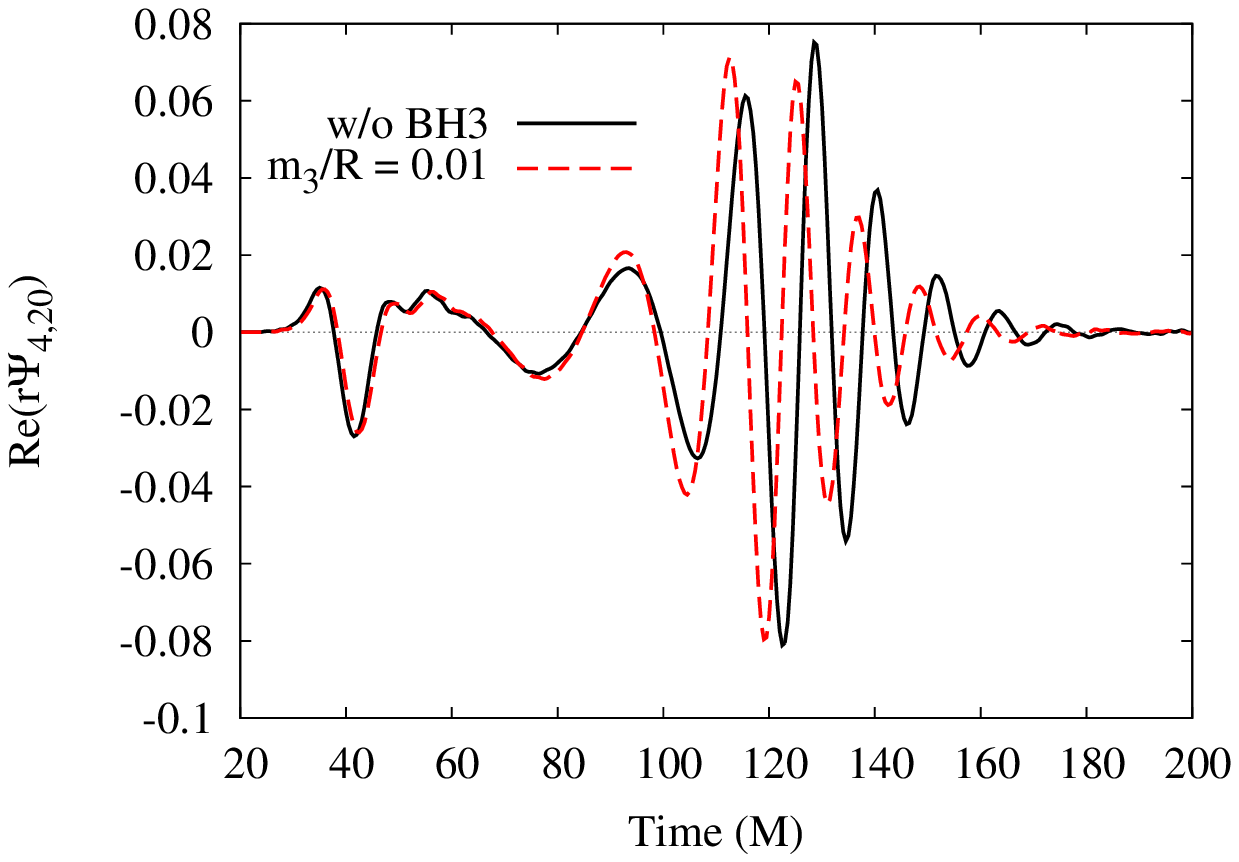}
\end{tabular}
\caption{Left panel: Typical inspiral trajectories for an isolated
inspiraling BBH and for the inspiraling-freefall case in
Sec.~\ref{infreefall}. The solid curve indicates the trajectory of
an isolated BBH and shows a quasi-circular inspiral shape. The (red)
dashed curve indicates the trajectory of the BBH in the
gravitational potential well of the third BH with $m_3=5\times10^4$
which is located at $(R=5\times10^6,0,0)$. There is a change in the
eccentricity of the (red) dashed curved when compared with the solid line.
Right panel: Real components of $r\Psi_{4,20}$ for an isolated
inspiraling BBH and for the inspiraling-freefall case. The red dashed
(black solid) line indicates the case with (without) the third BH.
The detector is located at $r=40$. It can be seen that the
gravitational potential of the third BH causes a tiny time delay in
the spurious part and the effect of a time advance in the major part
of the waveform.} \label{fig9}
\end{figure*}
%%%%%%%%%%%%%%%%%%%%%%%%%%%%%%%%%%%%%%%%%%%%%%%%%%%%%%%%%%%%%%%
%%%%%%%%%%%%%%%%%%%%%%%%%%%%%%%%%%%%%%%%%%%%%%%%%%%%%%%%%%%%%%%
\begin{figure}[tbp]
\centering\includegraphics[width=\columnwidth]{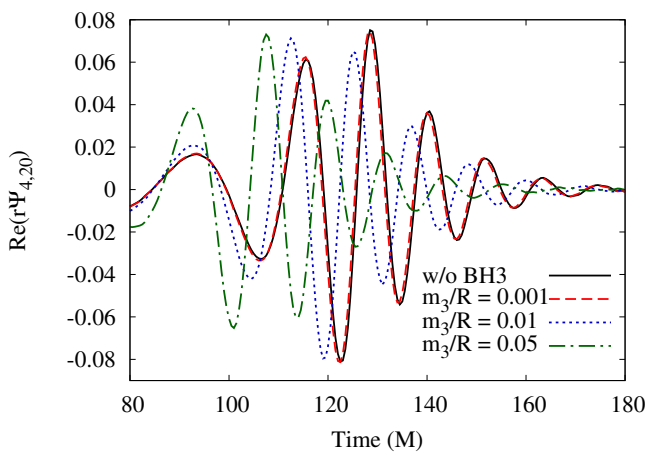}
\caption{The ${\rm Re}[r\Psi_{4,20}]$ for the inspiraling-freefall case
for different potential strengths of the third BH.
The results show the effect of advance in time and the prolongation of the
wavelength in the burst part due to the third BH's gravitational
potential. The effect of advance in time is obvious and  much larger than in
the spurious radiation (refer to Fig.~\ref{fig9}). 
The amplitude first increases and then decreases in the wave-train.
And the variation becomes bigger as the
gravitational potential of the third BH becomes larger.
This detail is explained in the context.}
\label{fig10}
\end{figure}
%%%%%%%%%%%%%%%%%%%%%%%%%%%%%%%%%%%%%%%%%%%%%%%%%%%%%%%%%%%%%%%
In this and the next subsections, we will consider the effect of a
third massive BH on an inspiraling BBH. The two BHs of the BBH
initially have the the irreducible masses $m_1=m_2\approx0.5$ and
are located at $(0,0,\pm\frac{D}{2})$. The third BH has mass
$m_3=5\times10^4$ and is located at $(R,0,0)$.

In this subsection, we study the inspiral-to-merger evolution of a
BBH while the CoM of the BBH freely falls towards the third large
BH. The left panel of Fig.~\ref{fig9} shows the evolution of a
typical inspiraling trajectory of the BBH, from the initial
coordinate separation $D=4.5$, affected by the third large BH
located at $(R=5\times10^6,0,0)$. We can see from the plot that the
gravitational potential of the third large BH leads to an increase
in the eccentricity of the trajectory of the BBH, compared with that
of an isolated BBH. The right panel of Fig.~\ref{fig9} shows a
typical comparison of the gravitational waveforms for the BBH
inspiraling with and without the influence of a third large BH. It
can be seen that the gravitational potential of the third large BH
causes a time advance, a prolongation of the wavelength, and a
somehow complicated variation in the amplitude of the gravitational
waveform.

As explained in Sec.~\ref{hoho}, the time delay for the existence of
a third large BH in a head-on collision arises from three factors:
the delayed merger process, the prolonged proper distance, and the
change of the coordinate time. For an inspiraling process of a BBH,
there exists another effect, i.e., a change of the eccentricity, as
shown in the left panel of Fig.~\ref{fig9}. It is obvious that the
change of the eccentricity is mainly caused by the tidal effect of
the third large BH. As a BBH system emits more gravitational
radiation in an elliptical orbit than in a circular orbit
\cite{peter64,linw90}, the increase of eccentricity is expected to
expedite the merger process of the BBH.
%\footnote{The effect of the gravitational circularization mentioned in Peters's
%work \cite{peter64} is under the assumption that the binary system is isolated.
%Therefore, the conclusion that the eccentricity of binaries tends to approach
%zero due to radiation reaction is not necessarily true and needs to be checked
%when the binary system is located in a gravitational potential from a third BH.
%This possibility has been pointed out in \cite{BPBMS09,IMFM10}. %}
Therefore, the time shift in the inspiraling process is the result of
competition between the increase in eccentricity and the others.
Here we look at this in more detail.

From the right panel of Fig.~\ref{fig9} we can see that the waveform is
affected by both the time delay and time advance phenomena: the spurious
radiation part of the waveform is delayed, while the burst part is
advanced. The beginning of the waveform, near $t=50$,
is a result of the spurious radiation hidden in the
initial data. Thus, this part is independent of the merging
process of a BBH. Here we can see a time delay because of the prolonged proper
distance from the BBH to the detector, at $r=40$, due to the existence of the
third BH.
In the main part of the merging process, starting at the coordinate time
$t\approx 100$, we can clearly see the time advance effect.
This is due to the increase in eccentricity under the
influence of the gravitational potential of the third large BH.
Obviously the effect of the eccentricity
increase overtakes other time delay factors in the competition.

Besides the effect of time advance, there is another feature shown
in the waveform. If we take the highest peak in the wave-train of
$\Psi_4$ roughly as the time of merger \cite{frans07}, this divides
the waveform in the gravitational wave-train into two kinds of
different behaviors: Before the merger, the wave amplitude of
$\Psi_4$ becomes larger as the gravitational potential of the third
large BH becomes larger. This is because the increase in
eccentricity of the BBH's trajectory is the dominant effect in
speeding up the inspiral-to-merger process compared to the isolated
case. This results in stronger gravitational radiation. After the
merger, the wave amplitude of $\Psi_4$ turns out to become smaller
as the gravitational potential of the third BH becomes larger. The
decrease in the amplitude of the gravitational waveform mainly
arises from the stronger redshift effect due to the existence of the
third BH. These phenomena are illustrated in the right panels of
Fig.~\ref{fig9} and in Fig.~\ref{fig10}.
%%%%%%%%%%%%%%%%%%%%%%%%%%%%%%%%%%%%%%%%%%%%%%%%%%%%%%%%%%%%%%%
\begin{figure}[tbp]
\centering\includegraphics[width=\columnwidth]{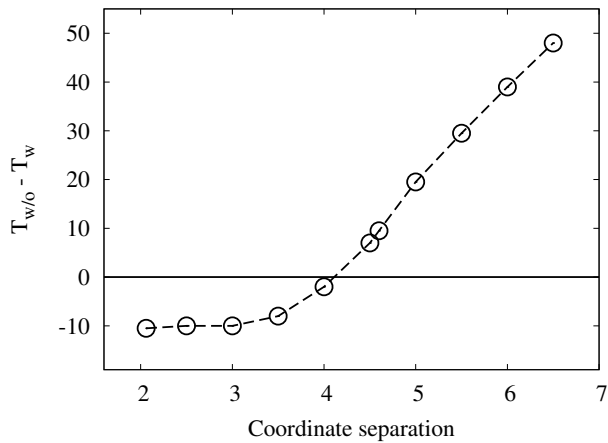}
\caption{Time difference of the inspiral-to-merger process between an isolated
BBH and the inspiral-freefall case in Sec.~\ref{infreefall}.
Here the merging time is the coordinate time corresponding to
the maximal amplitude of the $\Psi_{4,20}$ waveform.
The time difference ranges from being negative, i.e., time-delay, to being
positive, i.e., time-advance, as the initial separation increases.
In this case the turning point is around $r\approx 4.2$.}
\label{fig11}
\end{figure}
%%%%%%%%%%%%%%%%%%%%%%%%%%%%%%%%%%%%%%%%%%%%%%%%%%%%%%%%%%%%%%%

In fact, the inspiral-freefall case has been studied in
\cite{campanelli06} as an initial data problem. Based on the result
of its ISCO initial data, it was conjectured in the work that the
third BH will (1) increase the terminal amplitude of the inspiral
gravitational waveform; (2) increase the duration of the pre-plunge
phase; and (3) redshift the frequency of the gravitational wave
during the pre-plunge phase. Points (1) and (3) are consistent with
our current results, that the radiation power is amplified by the
third BH before the merging process, and that there is a
gravitational redshift effect. However, point (2) seems to conflict
with our time-advance result. Therefore, we need to look deeper
into this point. After a thorough study, we discover that as the
initial separation of BBH increases, the phase change of the
gravitational wave (due to the effect of the third large BH)
transits from a time-delay to a time-advance, as shown in
Fig.~\ref{fig11}. Note that the ISCO separation case studied in
\cite{campanelli06} corresponds to the leftest data point in this
figure, which is time-delay. One possible explanation for the result
is that, since the initial separation is small, the gravitational
redshift effect caused by the existence of a third large BH is
dominant in the evolution of the BBH and the merging time is
delayed. On the other hand, the eccentricity effect by
a third large BH outweighs the gravitational redshift effect when
the initial separation of the BBH becomes large enough, and thus the
merging time is advanced. In other words, the time difference in the
inspiral-to-merger process of a BBH under the influence of a third
large BH is dependent on the initial separation.

\begin{figure}[tbp]
\centering\includegraphics[width=\columnwidth]{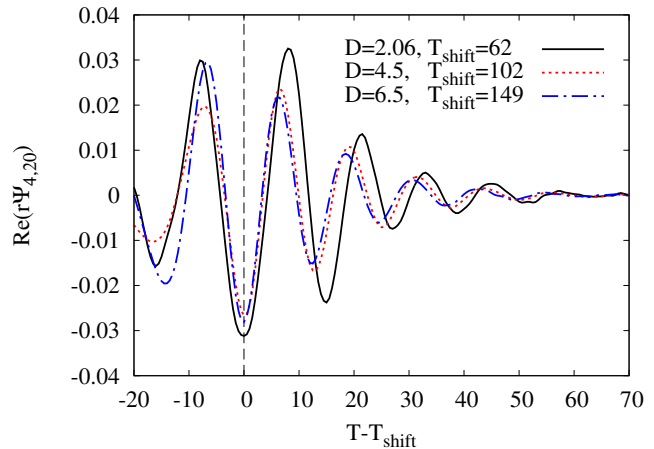}
\caption{Time-shifted waveforms of $r\Psi_{4,20}$ in the inspiral-to-merger
process of a BBH, with different initial separations, under the influence of
a third BH in the inspiral-freefall case. $T-T_{\rm shift}=0$ is the shifted
merging time defined at the maximal amplitude, and these waveforms do not match
with each other despite of the shifting.}
\label{redshiftvstidal}
\end{figure}

In Fig.~\ref{fig11}, We do not expect the time difference to converge when the
separation range is still very short, compared with a fixed $R$ (the distance
to the third BH).
Since the tidal effect on the evolution of the BBH due to the existence of the
third large BH is accumulative during the BBH's inspiral, we could expect that
the time-difference will keep increasing as the initial separation of the BBH
becomes larger \cite{CITE3}.
It is known that the gravitational waveform from the merger of a BBH with the
ISCO as the initial separation should agree with the waveforms just before the
merger from different initial separations.
However, this understanding is based on the assumption that the BBH is isolated.
It is not the case when there exists an external gravitational potential
background.
The final result should come from the competition between the circularization
effect from the gravitational radiation and the tidal effect from the external
gravitational potential background.
It seems not exist related investigations on this issue in the literature.
Therefore, we turn to check our numerical result.
In Fig.~\ref{redshiftvstidal}, it shows that the gravitational waveforms from
different initial separations between the two BHs in a BBH system under an
external gravitational potential do not match with one another even the
merging times have been shifted to be the same.
From our numerical experiments, it shows that the gravitational circularization
effect could not relax totally the accumulated eccentricity from the tidal
effect caused by the external gravitational potential background.
A detailed investigation is needed to have a better understanding on this
phenomenon in the future.
%%%%%%%%%%%%%%%%%%%%%%%%%%%%%%%%%%%%%%%%%%%%%%%%%%%%%%%%%%%%%%%
%%%%%%%%%%%%%%%%%%%%%%%%%%%%%%%%%%%%%%%%%%%%%%%%%%%%%%%%%%%%%%%

\begin{figure*}[tbp]
\begin{tabular}{rl}
\includegraphics[width=0.49\textwidth]{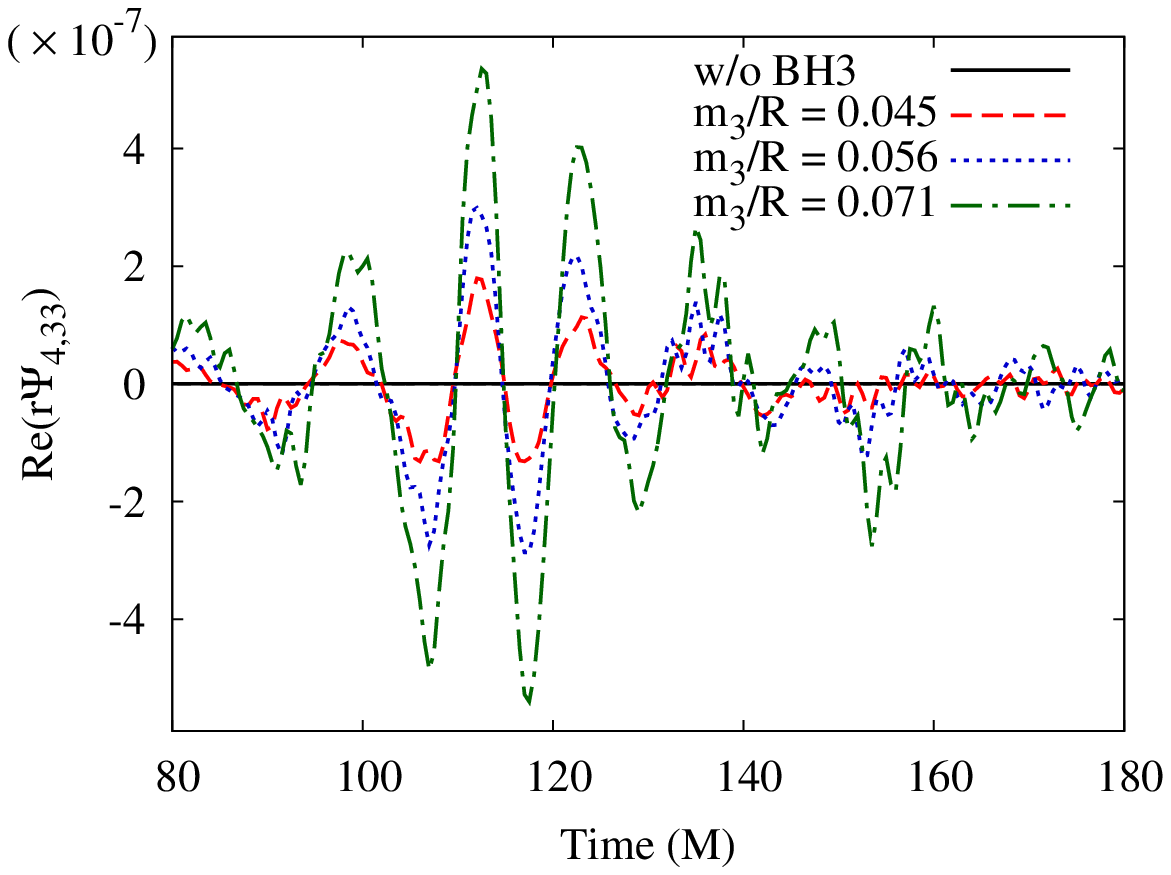} &
\includegraphics[width=0.49\textwidth]{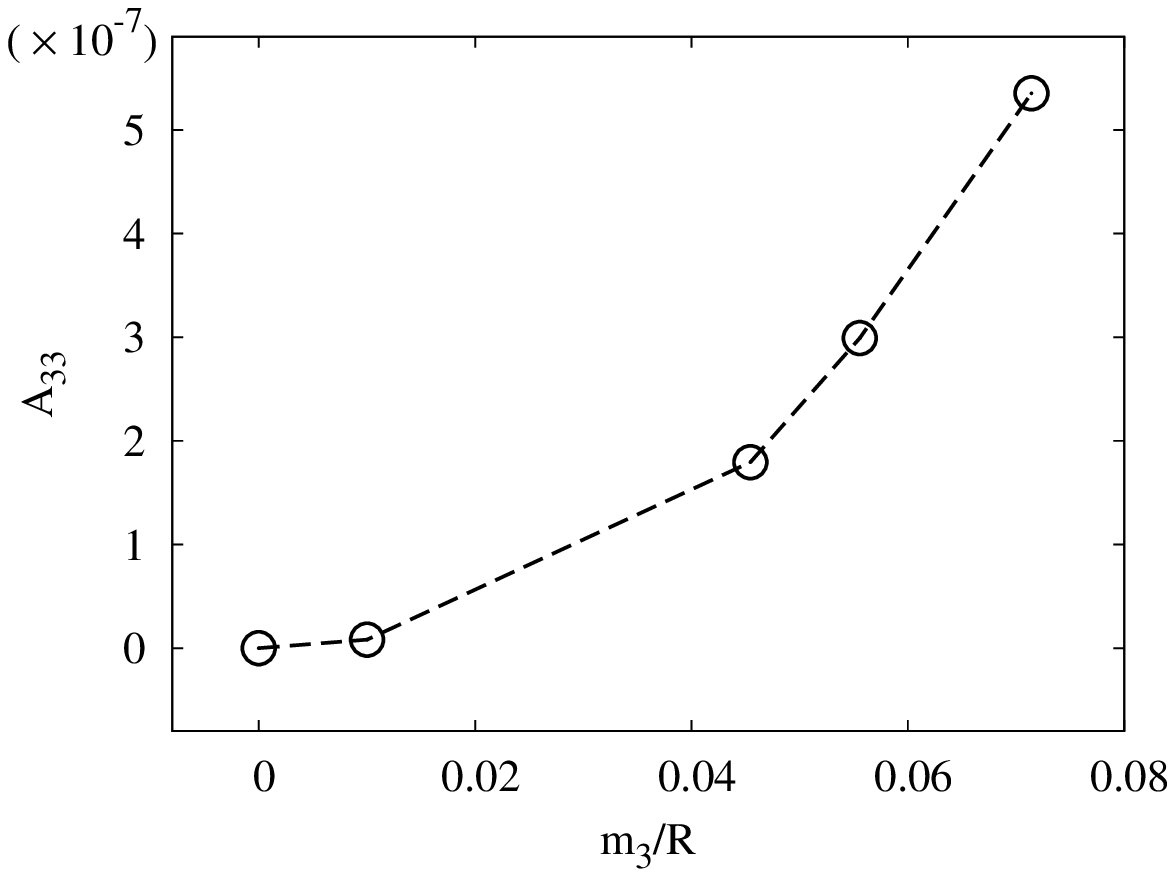}
\end{tabular}
\caption{Left panel: Comparison of ${\rm Re}(r\Psi_{4,33})$
in the inspiraling-freefall case with respect to different potential
strengths of a third large BH.
The existence of the third BH induces the higher-order ($\ell=3$)
mode and quite a large amount of the radiation is emitted away in this mode.
Right panel: Highest peaks of ${\rm Re}(r\Psi_{4,33})$ with respect to
different potential strengths of the third BH.
The nonlinear behavior of the higher-order ($\ell=3$) mode
with respect to the gravitational potential strength of the third BH is shown.}
\label{fig12}
\end{figure*}

Compared with the head-on collision of a BBH as discussed in Sec.~\ref{hoho},
the higher-order mode effect due to the existence of a third BH
in the inspiraling process is more evident. The left panel of
Fig.~\ref{fig12} shows the amplitude of ${\rm Re}(r\Psi_{4,33})$
produced during the inspiral-to-merger process of a BBH with respect to
different potential strengths of a third BH.
The higher-order mode, ${\rm Re}(r\Psi_{4,33})$, shows a nonlinear dependence
on the strength of the gravitational potential, as shown in the right panel of
Fig.~\ref{fig12}. This result is consistent with those detailed in the previous
two subsections: This nonlinear phenomenon could be a signature
distinguishing a three-BH system from a BBH system, as well as
complicating the identification of the source of a gravitational wave.
%%%%%%%%%%%%%%%%%%%%%%%%%%%%%%%%%%%%%%%%%%%%%%%%%%%%%%%%%%%%%%%
\subsection{Inspiraling-orbiting case}\label{inor}
%%%%%%%%%%%%%%%%%%%%%%%%%%%%%%%%%%%%%%%%%%%%%%%%%%%%%%%%%%%%%%%
%%%%%%%%%%%%%%%%%%%%%%%%%%%%%%%%%%%%%%%%%%%%%%%%%%%%%%%%%%%%%%%
%%%%%%%%%%%%%%%%%%%%%%%%%%%%%%%%%%%%%%%%%%%%%%%%%%%%%%%%%%%%%%%
%%%%%%%%%%%%%%%%%%%%%%%%%%%%%%%%%%%%%%%%%%%%%%%%%%%%%%%%%%%%%%%
\begin{figure*}[tbp]
\begin{tabular}{rl}
\includegraphics[width=0.49\textwidth]{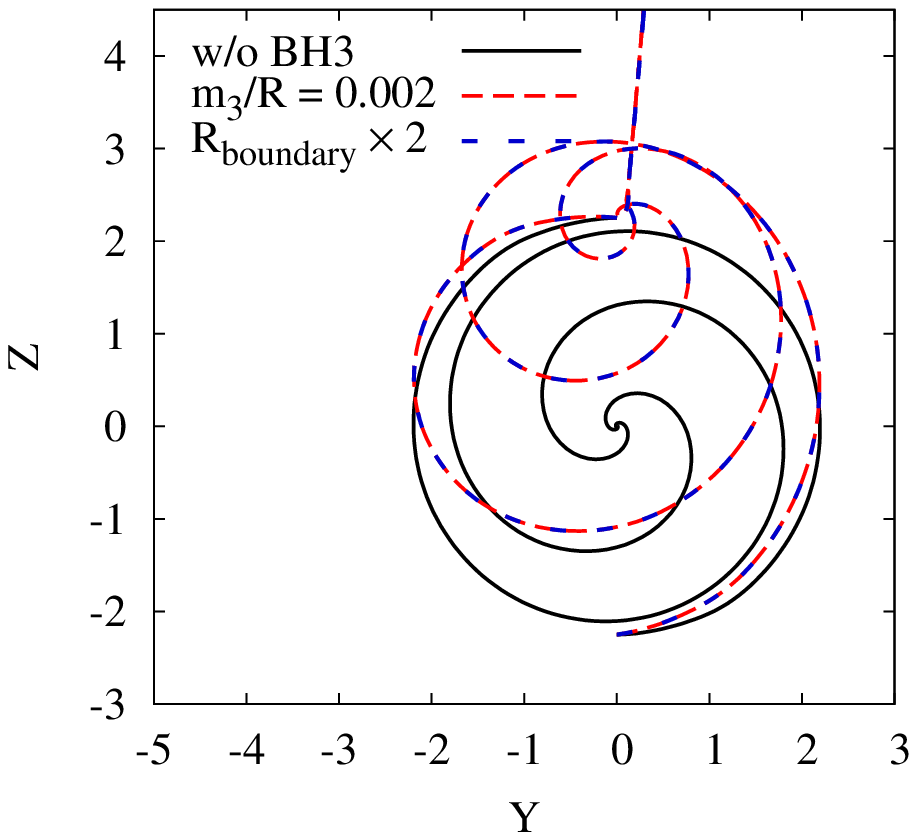}&
\includegraphics[width=0.49\textwidth]{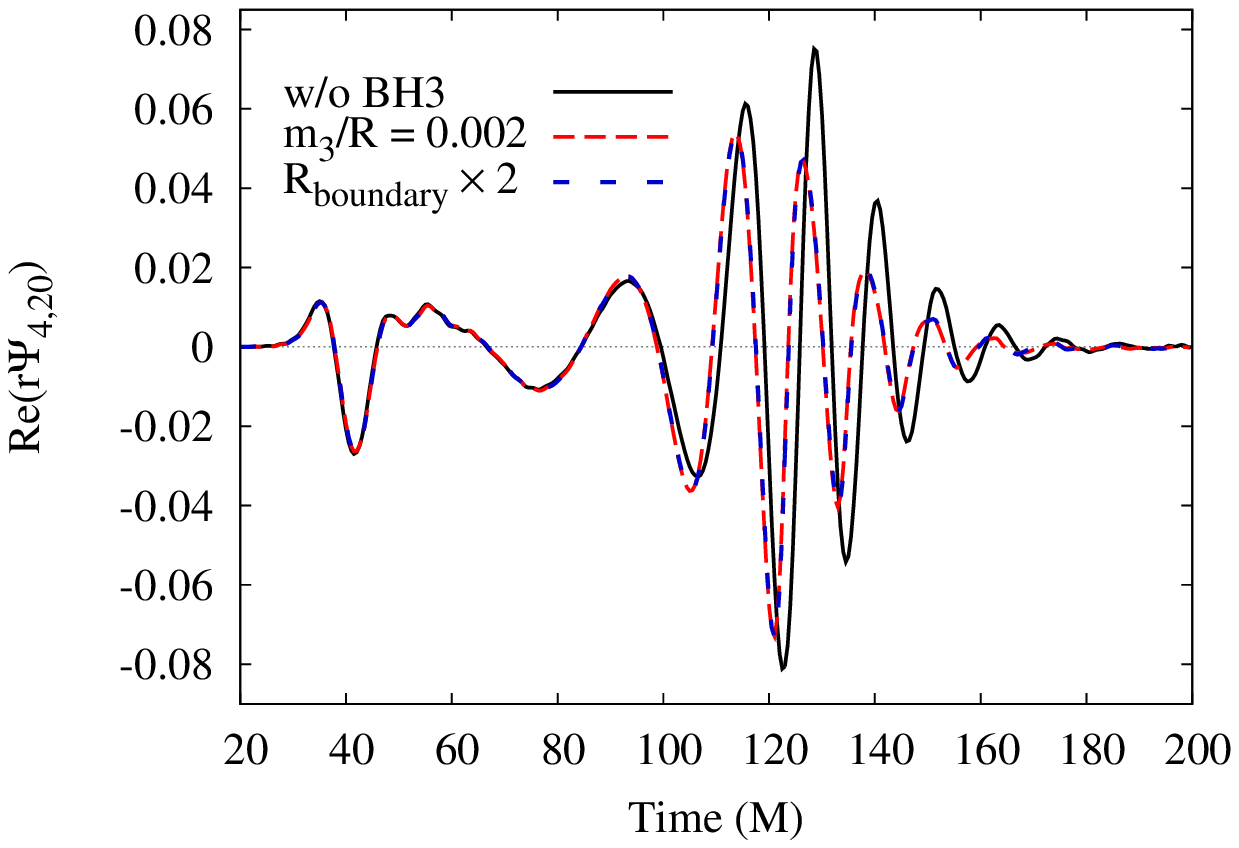}
\end{tabular}
\caption{Left panel: Typical trajectories for an isolated inspiraling
BBH and the inspiraling-orbiting case in Sec.~\ref{inor}. The
solid quasi-circular inspiral curve is for an isolated BBH. The red
dashed curve indicates the trajectory under the gravitational
potential of the third BH with $m_3=5\times10^4$ at
$(R=2.5\times10^7,0,0)$. The blue long-dashed line represents the
same case but with a doubled outer boundary. Right panel: The
corresponding waveforms of $r\Psi_{4,20}$. The red-dashed
(black-solid) line indicates the case with (without) the third
BH. The blue long-dashed line represents the case same as the one
for the red dashed curve but with a doubled outer boundary. It can
be seen that, besides the decrease in the wave amplitude, the
gravitational potential of the third BH causes a tiny time
delay in the spurious part and a time advance in the major part of
the waveform.} \label{fig13}
\end{figure*}
%%%%%%%%%%%%%%%%%%%%%%%%%%%%%%%%%%%%%%%%%%%%%%%%%%%%%%%%%%%%%%%
\begin{figure}[tbp]
\centering\includegraphics[width=\columnwidth]{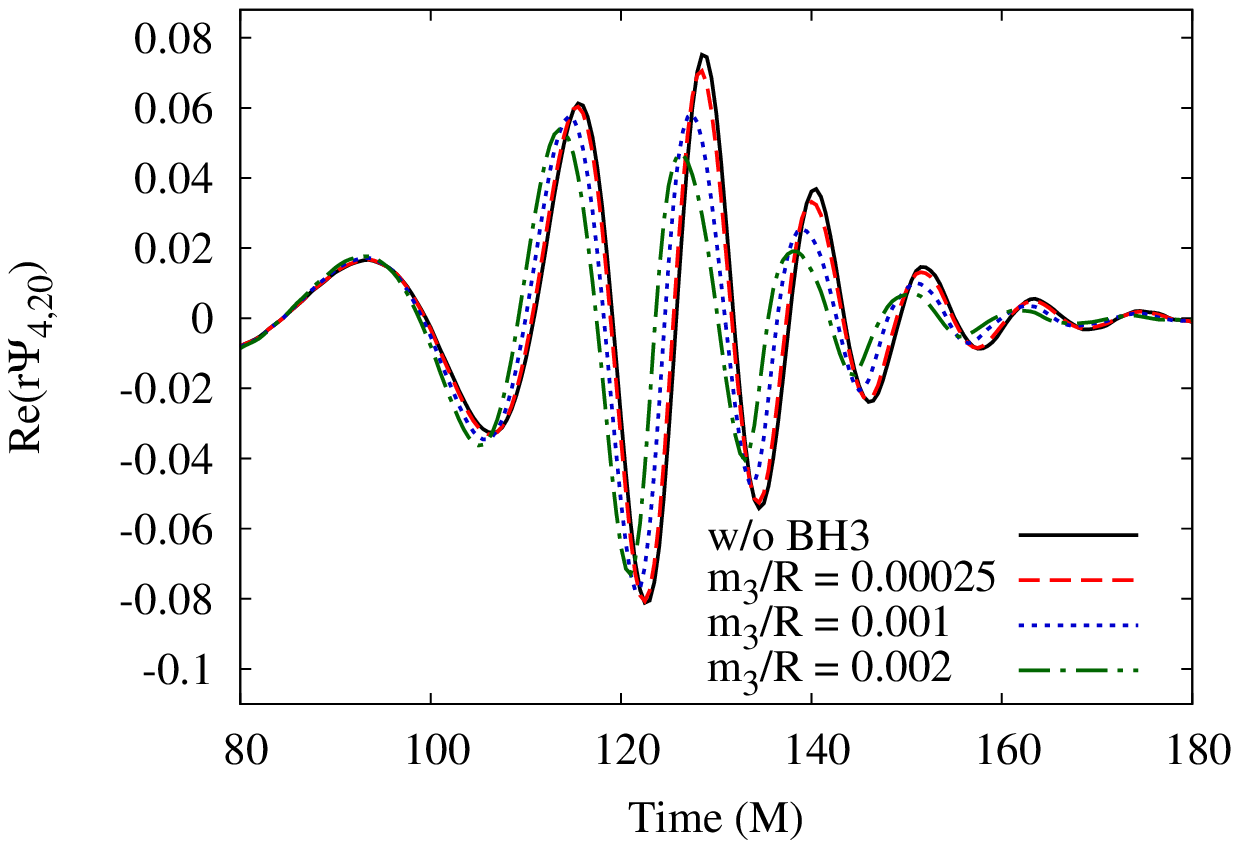}
\caption{The ${\rm Re}(r\Psi_{4,20})$
in the inspiraling-orbiting case with respect to
different potential strengths of the third BH.
Due to the existence of the third BH, the time advance effect
is clear and different from the spurious
radiation part, and the one for the inspiraling-freefall case (compare
Fig.~\ref{fig10}).
The amplitude decreases as the gravitational potential of the third BH
becomes stronger.}
\label{fig14}
\end{figure}
%%%%%%%%%%%%%%%%%%%%%%%%%%%%%%%%%%%%%%%%%%%%%%%%%%%%%%%%%%%%%%%
In this subsection, we investigate the inspiral-to-merger process of
a BBH orbiting around a third large BH. Initially, the CoM of the
BBH moves around the third large BH in a circular orbit. Therefore,
the velocity of the CoM of the BBH can be approximated by Newtonian
mechanics. Similar to the previous subsection, the two small BHs with the
irreducible mass $m_1=m_2\approx0.5$ are located at
$(0,0,\pm\frac{D}{2})$ initially. We put the third BH at $(R,0,0)$.
In order to get a circular orbit for the BBH around the third large
BH, we use the quantity $\sqrt{m_3/R}$ to evaluate the velocity of
the CoM of the BBH with respect to the third BH. This velocity is
then vectorially added to the quasi-circular orbiting velocity of each
BH of the BBH. We have ever tried measuring
the eccentricity of the small BBH affected by the background to
check if any extra eccentricity is introduced by the vectorial
addition. It turns out that there is no observable extra
eccentricity introduced by this addition during the whole inspiral
stage. For the third large BH, we fix its
mass parameter $m_3=5\times10^4$ as in the previous subsection while
varying its coordinate distance $R$. Given the mass parameters,
the linear momenta and the positions of the {\it three} BHs, the puncture
initial data was constructed as described in Sec.~\ref{NPuncture}.

The left panel of Fig.~\ref{fig13} shows a typical trajectory of the
BBH with the initial coordinate separation $D=4.5$ for the current
case, where the third large BH is located at $(R=2.5\times10^7,0,0)$.
It can be understood from the plot that the CoM of the BBH moves
roughly along a Newtonian circular orbit around the third large BH.
The trajectory is in fact the combination of the orbit of the CoM of
the BBH and the BBH's inspiraling around the CoM, and it is quite
deformed compared with the trajectory for an isolated BBH.

The gravitational waveform of the BBH in the current case is
presented in the right panel of Fig.~\ref{fig13}. Compared with the
right panel of Fig.~\ref{fig10}, we find that the effect of the
third large BH on the waveform is stronger in the current case than
in the inspiraling-freefall cases. In the plot, the waveform is
delayed in the spurious part but advanced in the major part,
caused by the gravitational potential of the third large BH. This is
similar to the previous inspiraling-freefall case. In
Fig.~\ref{fig13}, we also plot the trajectory and $\Psi_4$ for the
same scenario but with a doubled outer boundary. The perfect
coincidence of them indicates that the boundary condition described
in Sec.~\ref{secbc} is reliable.

In Fig.~\ref{fig14}, the time advance of ${\rm Re}(r\Psi_{4,20})$ due to
the third large BH is clear and different from its spurious radiation part.
This phenomenon is similar to the one in the inspiraling-freefall case
(compared with Fig.~\ref{fig10}).
Stronger gravitational potential of the third BH results in the smaller
amplitude of the waveform.
%%%%%%%%%%%%%%%%%%%%%%%%%%%%%%%%%%%%%%%%%%%%%%%%%%%%%%%%%%%%%%%
\begin{figure}[tbp]
\centering\includegraphics[width=\columnwidth]{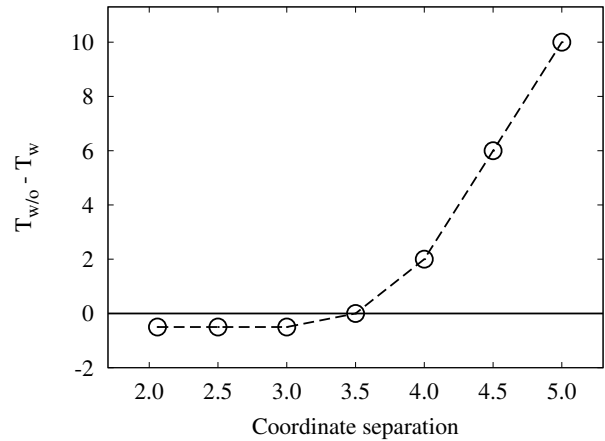}
\caption{Time difference of the inspiral-to-merger process between an
isolated BBH and the inspiraling-orbiting case in Sec.~\ref{inor}
with respect to the initial separation.
Here the merging time is the coordinate time corresponding to
the maximal amplitude of the $\Psi_{4,20}$ waveform.
The time difference ranges from being negative, i.e., time-delay,
to being positive, i.e., time advance, as the initial separation increases.
In this case, the turning point is around $r\approx 3.5$.}
\label{fig15}
\end{figure}
%%%%%%%%%%%%%%%%%%%%%%%%%%%%%%%%%%%%%%%%%%%%%%%%%%%%%%%%%%%%%%%
%%%%%%%%%%%%%%%%%%%%%%%%%%%%%%%%%%%%%%%%%%%%%%%%%%%%%%%%%%%%%%%
\begin{figure*}[tbp]
\begin{tabular}{cc}
\includegraphics[width=0.49\textwidth]{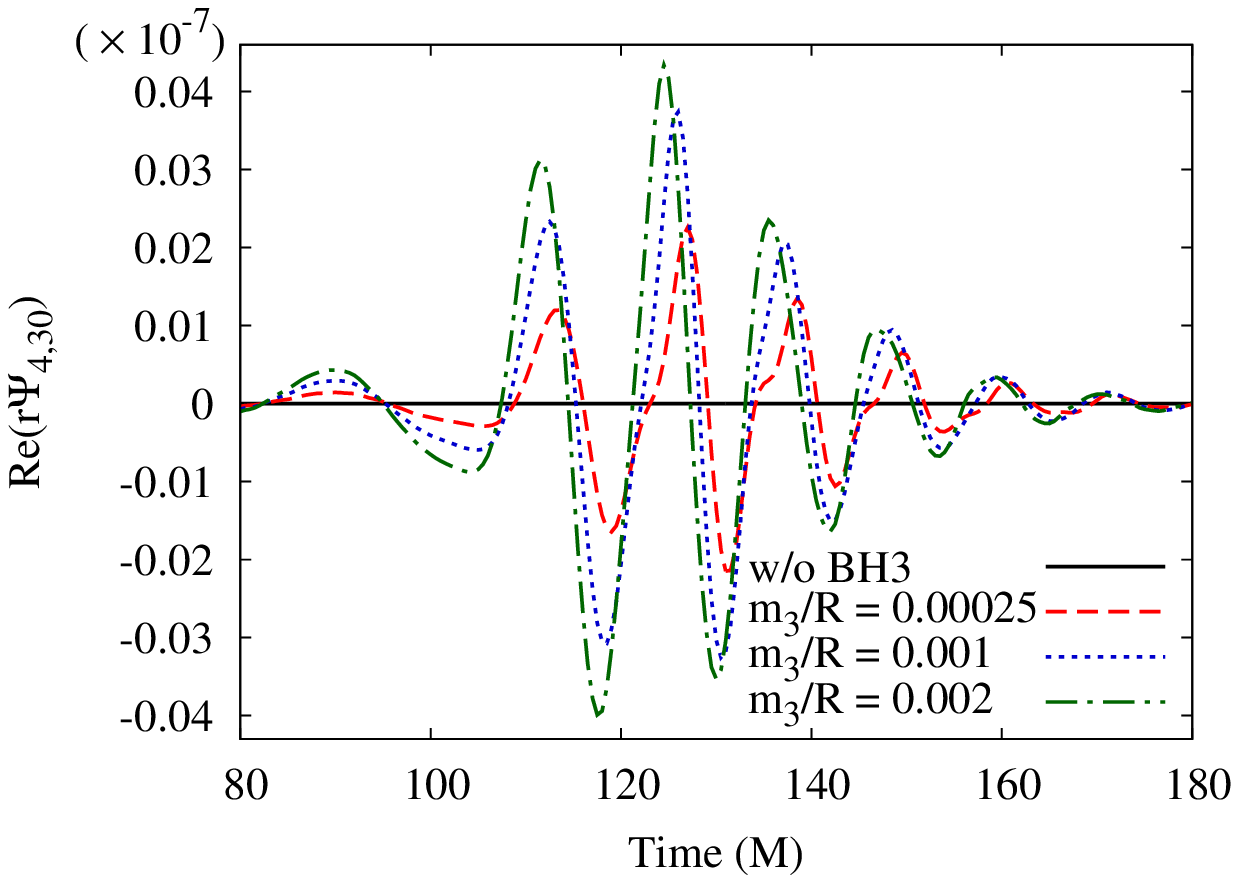} &
\includegraphics[width=0.49\textwidth]{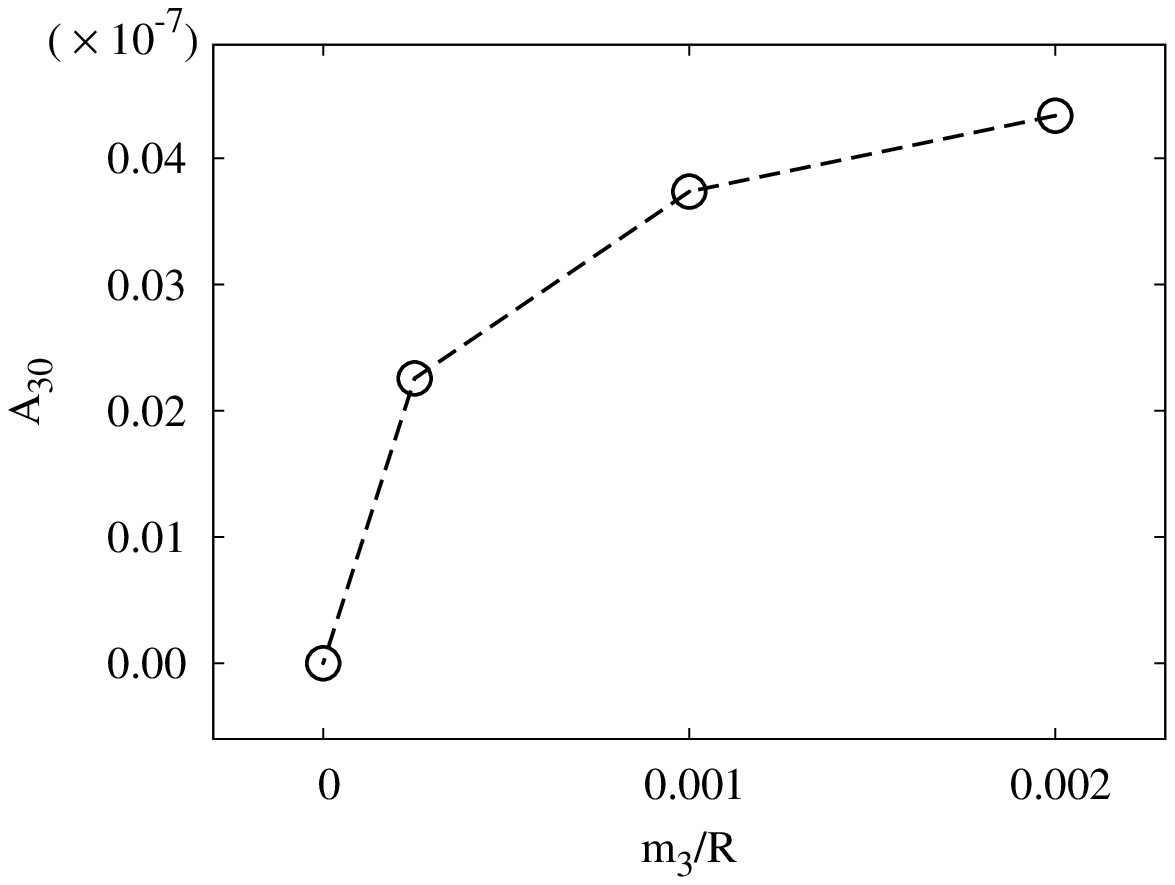}
\end{tabular}
\caption{Left panel: Comparison of ${\rm Re}(r\Psi_{4,30})$
in the inspiraling-orbiting case for different potential strengths of a third BH  (cf. Fig.~\ref{fig12}).
The amplitude is much larger than that in the inspiraling-freefall case.
Right panel: Highest peak with respect to different
potential strengths of the third BH. The nonlinear behavior of the
$\ell=3$ mode and the behavior of the curve are different from the one in
the inspiraling-freefall case.}
\label{fig16}
\end{figure*}
%%%%%%%%%%%%%%%%%%%%%%%%%%%%%%%%%%%%%%%%%%%%%%%%%%%%%%%%%%%%%%%

The shift of the waveform due to the gravitational background of the third BH
are quantitatively shown in Fig.~\ref{fig15}, in which it indicates the
transition
from the time-delay to time-advance as the initial separation of BBH increases.
Similar to the previous case,
the possible explanation for the change in the time difference of the
inspiral-to-merger process of a BBH under the influence of a third large BH
compared with the one of an isolated BBH, arises from the competition between
the eccentricity effect and the gravitational redshift effect.
Compared with the previous (inspiraling-freefall) case in which the
($\ell=m=3$) mode of $\Psi_4$ is dominant among the higher modes,
the ($\ell=3,m=0$) modes of $\Psi_4$ is dominant in the current case, as
shown in the left panel of Fig.~\ref{fig16}.
The orbiting of the BBH along the $x$-$z$ plane accounts for this effect.
In addition, the amplitude of the ($\ell=3,m=0$) mode of Re($\Psi_4$) increases
nonlinearly.
%%%%%%%%%%%%%%%%%%%%%%%%%%%%%%%%%%%%%%%%%
\section{Summary and discussion}\label{seciv}
%%%%%%%%%%%%%%%%%%%%%%%%%%%%%%%%%%%%%%%%%
Motivated by the fact that most BBH systems are located in
the gravitational potential of a super-massive BH hosted in the center
of a galaxy, we investigate the effect of the potential on the dynamics
of a BBH system.
Instead of the heavily numerical calculation of the evolution of the three-BH
problem with a fully relativistic treatment,
we use a perturbational scheme to investigate the effect of the gravitational
potential from a third large BH on the evolution of a BBH,
especially on the waveform of gravitational radiation.
In our perturbation method, we ignore the back-reaction of the BBH
system on the third large BH, regarding the third BH as
a background for the BBH system.

The scenarios we consider include the head-on collision and the
inspiral-to-merger process of a BBH in the cases of the BBH
system freefalling towards, or circularly orbiting around the third
large BH, which are considered as the two limits in all possible
configurations.
The effect of the gravitational potential from a third BH on a
BBH system in our study includes:
(1) the gravitational redshift effect, including the prolongation of proper
distance, the prolongation of the wavelength, the decrease of the waveform
amplitude;
(2) the increase in the eccentricity which is from the tidal effect of the
third large BH and expedites the merger of a BBH system;
(3) delaying or advancing the merging process of the BBH,
which depends on the competition between the gravitational redshift effect and
the eccentricity effect;
(4) inducing the higher-order modes in the gravitational waveform.
The orientation of the BBH system hardly affects the above results.
It is interesting that (4) supports the conjecture proposed in
\cite{torigoe09} in the full GR regime.
This might provide valuable information for gravitational wave astronomy to
distinguish whether the detected BBH system is isolated or located
in the potential of a super-massive BH.

In this work,
for the $\ell<3$ modes, all the quantities are linearly dependent on the
gravitational potential $m_3/R$, which validates the usage
of the perturbation method.
In most realistic cases, the gravitational potential is much weaker than
the cases considered in this work.
Therefore the perturbation method is expected to be applicable to most
realistic cases.

The third BH in our case introduces a non-axisymmetric background feature,
which makes the issue of gravitational wave extraction even more complicated. 
However, considering the detector located at the weak field region and  
finitely separated from both the merging binary and the massive BH, 
we adopt the usual approach to extract the information
of gravitational radiation via the Newman-Penrose Scalar $\Psi_4$.  
A possible extension of the work would be a detailed study on the
more rigorous way to extract the radiation under a background and the  
dependence of the modification on the relative position of the binary, detector
and the background source.

We have considered the evolution of an equal-mass BBH
under the influence of a third large BH. The study can be generalized
to an unequal-mass BBH system.
The major effects from the third large BH on a BBH are
shown qualitatively in this study.
The results indicate that these phenomena could introduce complications or
even be misleading in the
identification of the source of gravitational wave without further study to
distinguish the signatures of a BBH in a background
gravitational potential from the ones of an isolated BBH.

%%%%%%%%%%%%%%%%%%%%%%%%%%
%%%%%%%%%%%%%%
\section*{Acknowledgments}
%%%%%%%%%%%%%%%%%%%%%%%%%%%%%%%%%%%%%%%%
We thank F. Pretorius, H. Pfeiffer, E. Schnetter, and Y. Zlochower
for their useful discussion. This work was supported in part by the
National Science Council under the grants NSC98-2112-M-006-007-MY2 and
 NSC100-2112-M-006-005. Z.~Cao was supported by the NSFC
(No.~10731080 and No.~11005149). This work was also supported in
part by the National Center of Theoretical Sciences. We are grateful
to the National Center for High-performance Computing for the use of
their computer time and facilities. We are also grateful to the Academia Sinica
Computing Center for providing computing resource.
%%%%%%%%%%%%%%%%%%%%%%%%%%%%%%%%%%%%%%%%%%
%\section*{References}


\begin{thebibliography}{111}
\bibitem{CITE1}
The gravitational background effect depends in the perturbative
regime only on the ratio $m_3/R$. The ratio $m_3/R$ in our
simulation is in the range $m_3/R\in[0.001,0.1]$, which is much larger
than the ones in the usual astrophysical cases. Thus our result can
be applied straightforwardly to those usual cases. However, it is
quite possible for the existence of BBH systems near the center of a
galaxy where $m_3/R$ would be larger than in the usual astrophysical
cases by several orders of magnitude. Our study should be also applicable to
such cases as long as the ratio $m_3/R$'s are comparable.
\bibitem{CITE3}
The separation of the BBH can not increase
indefinitely under the assumption in this work. When the separation is
comparable with $R$, the whole configuration turns to a three-body problem 
to which the perturbational method in this work cannot apply.
\end{thebibliography}
\end{document}